\documentclass[a4paper,fleqn,usenatbib]{mnras}

\usepackage[T1]{fontenc}
\usepackage{ae,aecompl}

\usepackage{ulem}

\usepackage{graphicx}	
\usepackage{amsmath}	
\usepackage{amssymb}	

\newcommand{\msun}{\,\mathrm{M}_{\sun}}

\usepackage{natbib}
\usepackage{mathtools}

\usepackage[dvipsnames]{xcolor}

\def\gtsim {>\kern-1.2em\lower1.1ex\hbox{$\sim$}~}   
\def\ltsim {<\kern-1.2em\lower1.1ex\hbox{$\sim$}~}   

\usepackage{newtxtext,newtxmath}

\title[CNO dredge-up in APOGEE/Kepler red giants]{CNO dredge-up in a sample of APOGEE/Kepler red giants:
Tests of stellar models and Galactic evolutionary trends of N/O and C/N}

\author[F. Vincenzo et al.]{Fiorenzo Vincenzo$^{1}$\thanks{email: vincenzo.3@osu.edu} David H. Weinberg$^{1,2}$, Josefina Montalb\'{a}n$^{3}$, Andrea Miglio$^{4,5,3}$, 
\newauthor Saniya Khan$^{3}$, Emily J. Griffith$^{1}$, Sten Hasselquist$^{6,7}$, James W. Johnson$^{1}$, 
\newauthor Jennifer A. Johnson$^{1}$, Christian Nitschelm$^{8}$, Marc H. Pinsonneault$^{1}$
\\ ~ \\
$^{1}$Department of Astronomy \& Center for Cosmology and AstroParticle Physics, The Ohio State University, Columbus, OH 43210, USA \\
$^{2}$Institute for Advanced Study, Princeton, NJ 08540, USA \\
$^{3}$School of Physics and Astronomy, University of Birmingham, Edgbaston, B15 2TT, UK  \\ 
$^{4}$Dipartimento di Fisica e Astronomia, Universit\`{a} degli Studi di Bologna, Via Gobetti 93/2, I-40129 Bologna, Italy \\
$^{5}$INAF – Osservatorio di Astrofisica e Scienza dello Spazio di Bologna, Via Gobetti 93/3, I-40129 Bologna, Italy \\ 
$^{6}$Department of Physics \& Astronomy, University of Utah, Salt Lake City, UT, 84112, USA \\ 
$^{7}$NSF Astronomy and Astrophysics Postdoctoral Fellow \\
$^{8}$Centro de Astronom{\'i}a (CITEVA), Universidad de Antofagasta, Avenida Angamos 601, Antofagasta 1270300, Chile
}

\begin{document}

\pagerange{\pageref{firstpage}--\pageref{lastpage}} \pubyear{2021}

\maketitle

\label{firstpage}


\begin{abstract}
Surface abundances of C, N, and O in red giants are affected by processed 
material mixed into the stars' convective envelopes.  Using a sample of 
$\sim 5100$ stars with elemental abundances from APOGEE and asteroseismic
masses from {\it Kepler}, we test theoretical stellar models that predict this
mixing, then apply these models to derive birth C, N, and O abundances for 
these stars.  Our models with standard mixing can reproduce the observed trends
to within plausible uncertainties in the birth abundances.  Some models with
``extra'' mixing processes fail, predicting trends with surface gravity or
evolutionary state that are not observed.  Applying mixing corrections to the
APOGEE abundances removes the observed age-dependence of log(N/O) and log(C/N),
but it leaves trends of log(N/O) and log(C/N) with metallicity, as expected 
based on nucleosynthesis models.  The stellar N/O trend agrees well with Dopita
et al.'s calibration of gas phase log(N/O) with metallicity, and with gas phase 
trends in the MaNGA integral field survey of nearby galaxies.  We also find a 
substantial separation in birth [N/Mg] ratios between high-[$\alpha$/Fe]
(``thick disc'') stars and low-[$\alpha$/Fe] (``thin disc'') stars. We find a 
smaller but still clear separation for [C/Mg].  The trends of birth C and N
abundances with [Fe/H] and [$\alpha$/Fe] could affect spectroscopic age
estimates for red giants that rely on the observed C/N ratio as a diagnostic
of stellar mass.
\end{abstract}


\begin{keywords}
Galaxy: abundances -- stars: general -- stars: abundances -- stars: evolution -- stars: low-mass -- stars: atmospheres
\end{keywords}


\section{Introduction} \label{sec:intro}

The working assumption of the majority of chemical evolution studies is that present-day stellar photospheres retain the chemical composition of the molecular clouds from which stars were born. Since the stars within a given volume $\mathcal{V}(x,y,z)$ of the Milky Way (MW) are characterized by a distribution of ages, chemical evolution studies use atmospheric chemical abundances as measured in stars with different ages to constrain the chemical enrichment history of the interstellar medium (ISM) within $\mathcal{V}(x,y,z)$ from different nucleosynthetic sources (e.g., \citealt{romano2010,molla2015,prantzos2018,kobayashi2006,kobayashi2020}). 

One complication in interpreting such data is that radial migration of stars may mix populations with different enrichment histories, as pointed out by many observational (e.g., \citealt{hayden2015,casagrande2016,grieves2018,miglio2021}) and theoretical works (e.g., \citealt{schoenrich2009,loebman2011,bird2012,minchev2013,minchev2014,vincenzo2020,johnson2021}). A second complication, and the focus of this paper, is that internal mixing during stellar evolution causes the present-day atmospheric abundances in red giants to differ from their birth abundances, a phenomenon expected theoretically and demonstrated by observations in open cluster and globular clusters (e.g., \citealt{gilroy1989,korn2007,lind2008,souto2018,souto2019}). 

As stars ascend the red-giant branch (RGB), standard stellar evolution models predict C and N abundances to change at the stellar surface, because the convective envelope reaches increasingly deep regions inside the stars, dredging up material already processed by H-burning during the main sequence (MS). In this process (the so-called \textit{first dredge-up}), the pristine material in the stellar envelope is effectively mixed with processed material coming from the inner layers. The main effect of the first dredge-up is to increase the abundances of $^{4}$He, $^{13}$C, and $^{14}$N at the stellar surface, whereas $^{7}$Li and $^{12}$C are decreased (e.g., see the theoretical works of \citealt{iben1964,iben1967,deaborn1976,deaborn1978}, and the observational works of  \citealt{greene1969,lambert1977,kjaergaard1982,cottrell1986,shetrone1993,shetrone2019}). 

Because the reaction $^{14}$N(p,$\gamma$)$^{15}$O is much slower than other reactions in the CNO cycle, the abundance of $^{14}$N is drastically increased in regions of CNO burning, primarily at the expense of $^{12}$C.
In stars with increasing mass, the convective envelope during the RGB dredges up increasing amounts of CNO-processed material from regions where H was partly but not completely burned to helium on the MS. For this reason, the decrease of $\log(\text{C/N})$ and the increase of $\log(\text{N/O})$ in the stellar envelope at the end the first dredge-up correlate with the stellar mass, being larger for stars with higher mass. Because the age of a red giant is essentially determined by the MS lifetime, which is in turn determined by mass, many recent studies have investigated the use of $\log(\text{C/N})$ or spectral features that respond to C/N as diagnostics of stellar age, empirically calibrated using clusters, asteroseismology, or isochrones (e.g., \citealt{masseron2015,salaris2015,martig2016,ness2016,casali2019,miglio2021}).

By measuring the abundances of Li, C, N, and O in a sample of metal-poor red giants in the field, \citet{gratton2000} observed a significant depletion of C and Li as well as an enhancement of N after the bump in the RGB luminosity function (see Section \ref{sec:thermhaline} for more details and references). \citet{gratton2000} concluded that an \textit{extra-mixing process} is needed to explain the observed abundance changes in addition to standard mixing mechanisms such as the first dredge-up, which is experienced by red giants before they move across the RGB bump. Similar conclusions were also drawn by \citet{angelou2011,angelou2012,angelou2015}, who observed that the abundances of Li and C in a sample of RGB stars in metal-poor globular clusters are significantly depleted at luminosities above the RGB bump, whereas N is enhanced. 

Interestingly, by testing the predictions of models assuming extra-mixing due to thermohaline instability (see Section \ref{sec:thermhaline}), \citet{angelou2015} could reproduce the observed depletion of Li and C as a function of stellar luminosity above the RGB bump; however, fitting the observed relation of Li abundance vs stellar luminosity required a dimensionless parameter for the thermohaline mixing of Li that was a factor of $\approx5$ times lower than that required to fit the observed C vs stellar luminosity relation. More recently, \citet{lagarde2019} developed stellar population synthesis models showing that the assumption of thermohaline instability could help reproduce the trend of  $\log(\text{C/N})$ vs metallicity as observed in a sample of red giants in the field by the Gaia-ESO survey (GES; \citealt{gilmore2012,randich2013}), as well as the observed trend of $\log(\text{C/N})$ vs turn-off mass in a sample of red giants in open clusters and globular clusters. 

Understanding CNO evolution is important for testing stellar astrophysics, for interpreting results of Galactic chemical evolution surveys that target red giants, and also for extragalactic studies. 
Nebular N and O abundances can be measured from the analysis of the emission lines in the spectra of individual HII regions within our Galaxy (e.g., \citealt{shaver1983,vilchez1996,afflerbach1997,rudolph2006,fernandez-martin2017,esteban2018}) and extragalactic systems (e.g., \citealt{pagel1981,garnett1987,berg2020,esteban2020}). Statistical samples of N and O abundance measurements in extragalactic systems are also available thanks to large spectroscopic surveys of galaxies in the nearby Universe (e.g. \citealt{andrews2013,perez-montero2016,belfiore2017,schaefer2020}). Examples of such surveys are  Mapping Nearby Galaxies at Apache Point Observatory (MaNGA; \citealt{bundy2015}) and Calar Alto Legacy Integral Field Area (CALIFA; \citealt{sanchez2012}). 
Many studies of the gas-phase metallicities in these 
integral field surveys and integrated galaxy spectra at low and high redshift make use of strong-line diagnostics based on N/O (e.g., \citealt{kewley2002,pettini2004,perez-montero2009,dopita2016}). This choice is justified by the fact that the nucleosynthesis of N in asymptotic giant-branch (AGB) stars and massive stars strongly depends on metallicity (see \citealt{henry2000,vincenzo2016}, and references therein), and a tight correlation between N/O and O/H in the ISM of galaxies is observed by several studies (e.g., see the recent work of \citealt{berg2020}, and references therein), as well as predicted by cosmological hydrodynamical simulations (e.g., \citealt{vincenzo2018b}). By analysing a sample of star-forming galaxies observed by MaNGA, \citet{schaefer2020} found that the observed variation of N/O at fixed O/H correlates with the local star formation efficiency, in agreement with the predictions of chemical evolution models \citep{vincenzo2016}. 

In this work, we study how C, N, and O abundances vary in a sample of $\sim5100$ red giants observed by the the Apache Point Observatory Galactic Evolution Experiment (APOGEE; \citealt{majewski2017}) and the \textit{Kepler} mission \citep{borucki2010}. From the asteroseismic analysis of the \textit{Kepler} light curves of these stars, \citet{miglio2021} could accurately constrain their masses and radii. \cite{pinsonneault2018} carried out a similar effort with
APOGEE+{\it Kepler} stars using different methodology for the
asteroseismic modeling. Our paper has three inter-locking goals. First, we use the APOGEE+\textit{Kepler} observations to test the predictions of different  stellar  evolution models, including either standard mixing predictions or extra-mixing mechanisms (see Section \ref{sec:test}). Second, we apply the theoretical mixing corrections to infer the birth C, N, and O abundances of our APOGEE sample, which we use to examine trends of elemental abundance ratios as a diagnostic of chemical enrichment scenarios (see Section \ref{sec:correcting}). Third, recognizing the large uncertainties in empirical metallicity calibrations based on strong emission lines (e.g., \citealt{kewley2008}), we compare the evolution-corrected N/O vs O/H trend in our red giant sample to gaseous nebulae in the Local Universe to constrain the best metallicity calibration (see Section \ref{sec:comparing}). 

\section{Data and Models} 
\label{sec:model}

The surface abundances of N and C (and to some extent O) in evolved stars
differ from their birth abundances because stellar material that has been
processed by CNO burning is mixed into the convective envelope.  The ratio
N/C increases drastically in regions where CNO burning has reset abundances
to the equilibrium determined by nuclear reaction rates.  In solar core
conditions, for example, the birth ratio ${\rm N}/{\rm C} \approx 0.25$
is reset to ${\rm N}/{\rm C} \approx 225$.  Surface abundances of N and C
therefore depend on the fraction of material now in the convective envelope
that was once in a zone subject to the CNO cycle.  Typically O abundances are little
affected because the branch of the CNO cycle involving $^{16}$O only
comes into play at high temperatures.  We consider several different
models for mixing in RGB and RC stars, as a function of mass and metallicity.
To test these models against data and derive evolution-corrected
abundances, we require stars with measured C, N, O surface abundances,
overall metallicity, and masses.  For this purpose, we use
Miglio et al.'s (\citeyear{miglio2021}) sample of APOGEE stars with mass
estimates based on {\it Kepler} asteroseismology. In our analysis, chemical abundance measurements are from APOGEE-DR16 \citep{wilson2019,ahumada2020,jonsoon2020}, which are scaled with respect to the solar abundances as measured by \citet{grevesse2007}. 

\subsection{The sample of red giants of \citet{miglio2021}}
\label{sec:miglio-sample}

Using a combination of spectroscopic and asteroseismic constraints \citet{miglio2021} inferred masses, radii, and ages of about 5400\ red-giant stars observed by {\it Kepler} and APOGEE. In this work we adopt stellar parameters resulting from their reference run R1, and refer the reader to their Table 1 and Sec. 2 for a discussion of the systematics that affect the accuracy of the inferred properties, both related to biases in the data and from the grid of stellar models adopted in the modelling code.  

From the initial sample of giants we exclude  stars with an inferred radius larger than 30 R$_{\odot}$ to avoid stars in the poorly-tested low-frequency domain, thus adopting less restrictive criteria than those used in Sec. 4 of \citet{miglio2021}. We obtain a sample of $\sim5100$ stars with a median random uncertainty of  $6 \%$ in mass and of $25\%$ in age.

\subsection{The MESA stellar evolution models}
\label{sec:MESA-models}

This grid of stellar models has been computed using the code {\tt MESA} \citep[version 11532]{paxton2011,paxton2013,paxton2015,paxton2018}\footnote{The code {\tt MESA} and the relevant documentation are publicly available at the following link: \url{http://mesa.sourceforge.net}.}. It covers masses from 0.6 to 3.0$\msun$ with a step of 0.05$\msun$ and evolution phases from zero age main sequence (ZAMS) to the first AGB-pulse. Our solar-mixture reference is that from \citet[AGSS09]{asplund2009}, and in addition to the solar scaled chemical composition, we also consider $\alpha$-element enhancement of 0.2 and 0.4 dex\footnote{In the MESA stellar evolution models, the $\alpha$-elements include the following chemical elements: O, Ne, Mg, Si, S, Ar, Ca and Ti.}. For each of these reference mixtures we consider eight different values of [Fe/H], ranging from -1.50 to +0.25. Radiative opacity values for these specific chemical element mixtures come from OPAL opacity tables \citep{iglesias1996} complemented with tables by the Wichita State University for the low-temperature domain \citep{ferguson2005}. Nuclear reaction rates are from the NACRE compilation \citep{angulo1999}, and the equation of state is computed with the code {\tt FreeEOS} \citep{irwin2012}.  Concerning transport process, we adopt the mixing-length formalism of convection \citep{bohm1958} with a mixing-length parameter derived from solar calibration with the same physics, which is kept fixed for the whole grid. These models also include the effects of gravitational settling following the approach by \citet{thoul1994}, with the efficiency of microscopic diffusion slightly decreased by including a turbulent diffusion term \citep{morelthevenin}.  

Concerning chemical mixture at the border of convection regions, we do not consider extra-mixing at the limit of the convective core in H-burning MS stars. Assuming a diffusive convective core overshooting has a negligible effect on the predictions of our models for stars with initial mass $m \lesssim 2 \,\text{M}_{\sun}$; for stars with mass $m \gtrsim 2\;\text{M}_{\sun}$, increasing the convective core overshooting efficiency, $f_{\text{ov,core}}$, from $0$ to $0.02$ causes a further enhancement of the predicted N abundances at the end of the first dredge up by $\approx0.04\,\text{dex}$. 
For the central He-burning phase, we adopt the \citet{bossini2015} formalism, in which extra-mixing is allowed at the limit defined by Schwarzchild criterion and whose extension has been checked against the AGB-bump location. Finally, we also assume a diffusive extra-mixing at the bottom of the convective envelope, following the parametrisation by \citet{herwig1997},  with an overshooting parameter $f_{\text{ov}}=0.02$ \citep{khan2018}. \citet{khan2018} found that models with higher $f_{\text{ov}}$ at low metallicities could better reproduce the position of the RGB bump; in this paper, we do not modify the value of $f_{\text{ov}}$ for different metallicity models. We find that increasing $f_{\text{ov}}$ from $0.02$ to $0.05$ causes a further increase of $\log(\text{N/H})$ at the end of the first dredge up by $\approx0.03$ in stars with mass $m = 1\,\text{M}_{\sun}$, regardless of their metallicity, whereas it does not affect the predicted surface abundances of stars with mass $m>1\,\text{M}_{\sun}$.

\subsection{The stellar evolution models of \citet{lagarde2012} with rotation-induced mixing and thermohaline instability}
\label{sec:lagarde-models}

\subsubsection{Rotation-induced mixing}

If stars have non-negligible rotation on the zero-age main sequence, the dynamical evolution of their angular momentum in the radiative regions is governed by an advection-diffusion partial differential equation (e.g., see \citealt{zahn1992,maeder1998}, and equation 4 of \citealt{lagarde2012}); the spatial derivatives in this equation are associated to angular momentum transport mechanisms, which correspond to \textit{(i)} meridional circulation (advection term, inducing differential rotation) and \textit{(ii)} shear-induced turbulence (diffusion term, inducing uniform rotation). 

The meridional circulation and shear turbulence mechanisms governing the angular momentum transport induced by rotation cause - together with microscopic diffusion -- a vertical transport of chemical elements in the radiative regions of the stars \citep{chaboyer1992,charbonnel1992,chaboyer1995,decressin2009b}
which can be  treated as a diffusion process, simplifying the equations governing the dynamical evolution of the concentration of the chemical elements in stellar evolution models (e.g., see \citealt{chaboyer1992,charbonnel1995}, and equation 5 of \citealt{lagarde2012}). 

\subsubsection{Thermohaline instability}
\label{sec:thermhaline}

Another proposed mixing process in the radiative regions of red giants -- which can effectively alter their surface abundance composition -- is the thermohaline instability, which is a double-diffusive process (see \citealt{eggleton2006,charbonnel2007,charbonnel2010,cantiello2010}, and the recent review of \citealt{garaud2021}). Thermohaline instability is predicted to take place after the bump in the luminosity function of RGB stars \citep{thomas1967,iben1968}. In summary, the RGB bump happens after the first dredge-up has reached its deepest extent and the hydrogen-burning shell advances in mass until it interacts with the H abundance discontinuity, which has been left behind by the withdrawing convective envelope, at the interface between the radiative region and the deepest point of the convective region. When reaching the discontinuity, the mean molecular weight, $\mu$, decreases, causing a sudden decrease of the stellar luminosity for a short period ($L\propto\mu^{7.5}$; e.g., see \citealt{schwarzschild1958}).

After the bump in the luminosity function, the hydrogen burning shell advances further through a region which has been homogenized by the convective envelope, where there is a gradient 
$\text{d}\ln\mu/\text{d}\ln P$ in the mean molecular weight, $\mu$, with respect to the pressure, $P$. It is at this time that thermohaline instability is predicted to take place in the radiative region, being induced by an inversion of $\text{d}\ln\mu/\text{d}\ln P$, due to the thermonuclear reaction $^{3}\text{He}( ^{3}\text{He}, 2\text{p}) ^{4}\text{He}$ in the outermost layers of the H-burning shell. This reaction can lower the mean molecular weight in a medium which has been chemically homogenized by the first dredge-up, creating lighter fluid parcels which can quickly diffuse to the surface. 

Thermohaline mixing is included in stellar evolution models by adding an additional diffusion term in the equation for the dynamical evolution of the concentration of the chemical elements (e.g., see equation 5 of \citealt{lagarde2012}). Because of the difficulty of developing high-resolution 3-D models to study in detail this double-diffusive mechanism in stars, the formulation adopted by stellar evolution models like those of \citet{lagarde2012} leads to a mixing efficiency parameter for thermohaline instability that has been disputed by other numerical experiments based on 2-D models, which find an equivalent efficiency parameter $\approx 10$ times lower (e.g., see \citealt{denissenkov2010,denissenkov2011}, and the discussions in \citealt{angelou2015,shetrone2019}, and Section 3.6 of \citealt{garaud2021}). 

\subsubsection{The models of \citet{lagarde2012}}

To test the impact of rotation-induced mixing and thermohaline instability on the evolution of the atmospheric chemical abundances in the red giants of our sample, we analyse the predictions of the grid of models as developed by \citet{lagarde2012} by making use of the Grenoble stellar evolution code ({\tt STAREVOL}; see \citealt{siess2000,palacios2003,palacios2006,decressin2009}).  \citet{lagarde2012} also provide a grid of stellar models with standard mixing prescriptions (i.e., no mixing processes are assumed to take place outside of the convective regions), which are used to validate the predictions of the models with extra-mixing prescriptions. 

In our analysis, we consider both the models with extra-mixing processes and the models with standard mixing prescriptions. Both grids are computed for initial masses in the range $0.85 \le m \le 6\,\text{M}_{\sun}$ and the following metallicities: $Z=10^{-4}$, $2 \times 10^{-3}$, $4 \times 10^{-3}$, and $0.014$\footnote{The grids of stellar models of \citet{lagarde2012} can be downloaded at the following link: \url{https://vizier.u-strasbg.fr/viz-bin/VizieR?-source=J/A+A/543/A108}.}.

\begin{figure}    
\centering
\includegraphics[width=8cm]{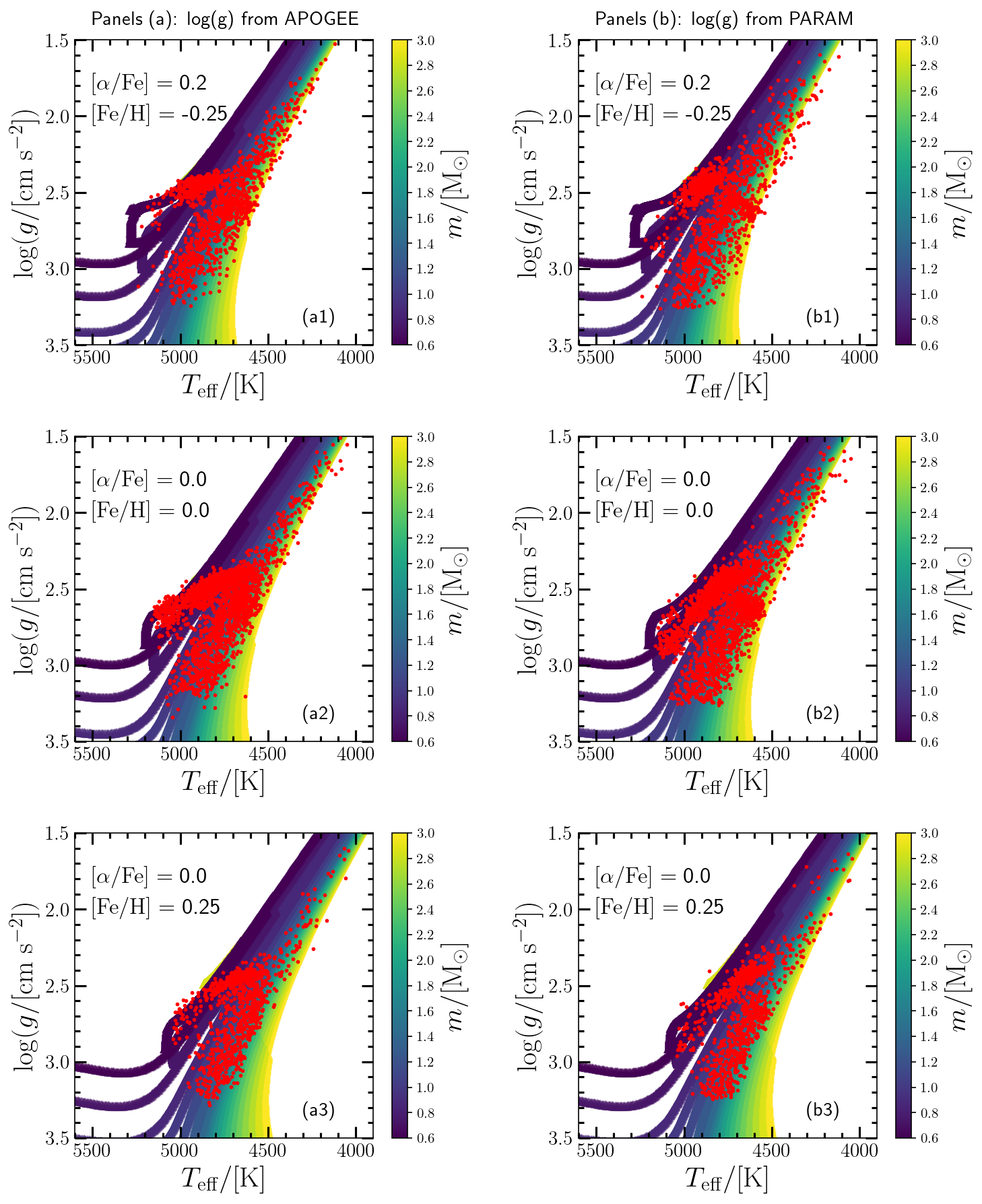}
\caption{The Kiel diagram with the red giants in the sample of \citet[red points]{miglio2021} as compared with the predictions of our reference MESA stellar evolution models for different stellar masses (color-coding). The stellar models have the $[\alpha/\text{Fe}]$ and [Fe/H] abundances reported in the top left corner of each panel. From top to bottom, stars are shown in different bins of $\text{[Fe/H]}$ with width $\Delta\text{[Fe/H]}=0.25$, centered at the value reported in the top left corner. To understand the impact of the systematics in the measurement of $\log(g)$ with different methods, in panels (a1-a3) we use the spectroscopic $\log(g)$ values from APOGEE-DR16, whereas in panels (b1-b3) we use the asteroseismic $\log(g)$ values. 
}
\label{fig:kiel-diagram}
\end{figure}

\begin{figure*}    
\centering
\includegraphics[width=16cm]{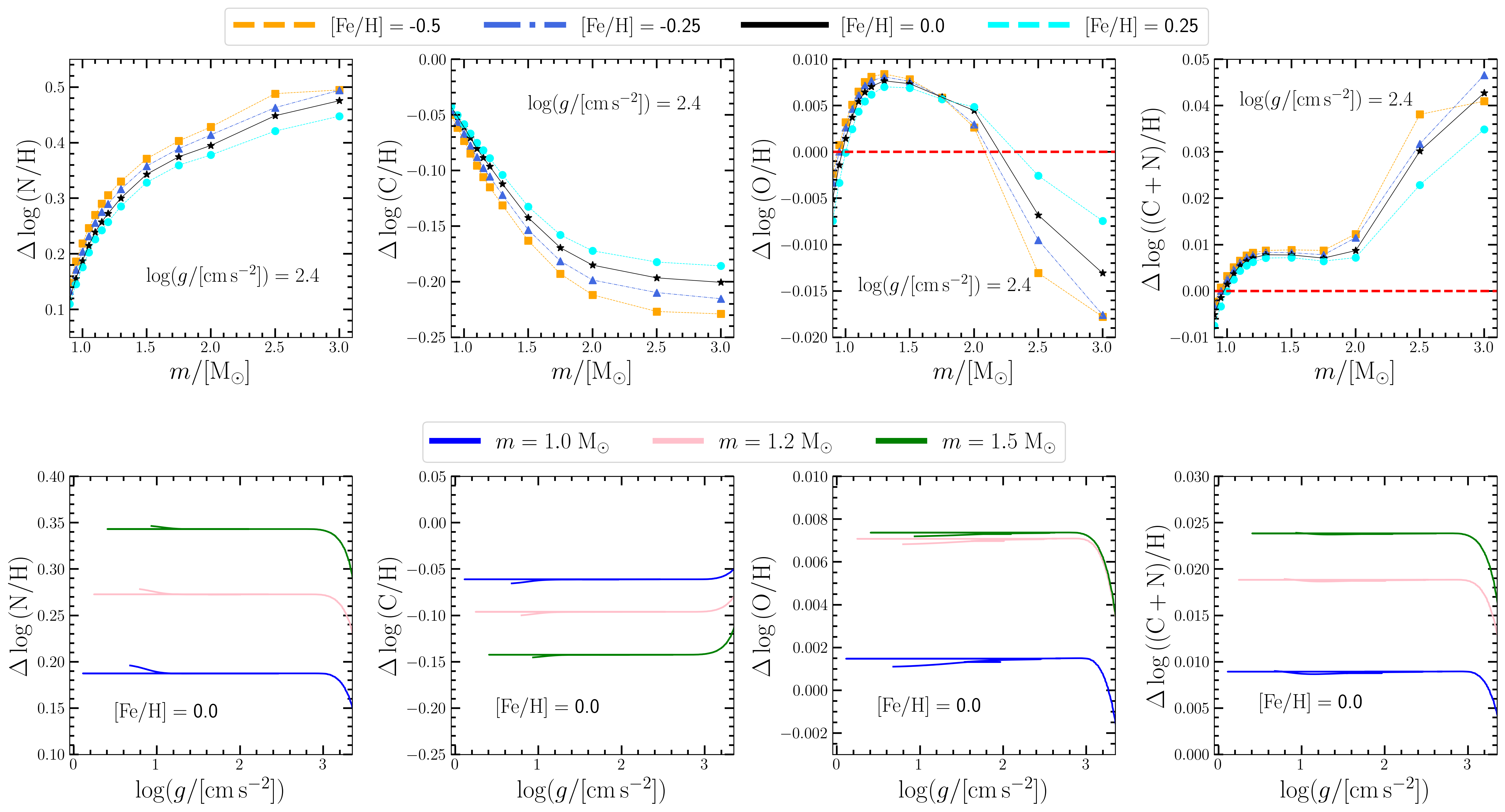}
\caption{The surface abundance variations $\Delta\log(\text{N/H})$, $\Delta\log(\text{C/H})$, $\Delta\log(\text{O/H})$, $\Delta\log(\text{(C+N)/H})$ as predicted by our reference MESA stellar evolution models with respect to the assumed birth abundances. The top panels show the predicted trends vs stellar mass, for different metallicities (orange dashed curves and squares: $[\text{Fe/H}]=-0.5$; blue dashed-dotted lines and triangles: $[\text{Fe/H}]=-0.25$; black solid line and star symbols: $[\text{Fe/H}]=0$; cyan dashed lines and circles: $[\text{Fe/H}]=0.25$), whereas the bottom panels show the predicted trends vs $\log(g)$ for different assumed stellar masses (blue solid lines: $m=1\,\text{M}_{\odot}$; pink solid lines: $m=1.2\,\text{M}_{\odot}$; green solid lines: $m=1.5\,\text{M}_{\odot}$). }
\label{fig:deltaCNO_models}
\end{figure*}

\section{Tests of evolutionary mixing models}
\label{sec:test}

Fig.~\ref{fig:kiel-diagram} shows the Kiel diagram with the red giants in the sample of \citet[red points]{miglio2021}. The different panels contain stars in different ranges of [Fe/H] abundances, centered -- from top to bottom -- at $[\text{Fe/H}]=-0.25$, $0.0$, and $0.25\,\text{dex}$. To understand the impact of the systematics in the measurement of $\log(g)$ with different methods, the stars in the Kiel diagrams in Fig. \ref{fig:kiel-diagram}(a1-a3) have spetroscopic $\log(g)$ values measured by APOGEE-DR16, whereas the stars in Fig.~\ref{fig:kiel-diagram}(b1-b3) have asteroseismic $\log(g)$ values that we computed from the stellar radii and masses as derived by {\tt PARAM}  \citep{dasilva2006,rodrigues2014,rodrigues2017}\footnote{The web interface of {\tt PARAM} with the relevant documentation is available at the following link: \url{http://stev.oapd.inaf.it/cgi-bin/param}.}. The main effects of asteroseismic vs spectroscopic $\log(g)$ values are in the apparent morphology and $\log(g)$ distribution of RC stars, as well as in the spread at low $\log(g)$ above the RC, in the region populated by upper RGB and AGB stars. Since most of the red giants in our sample are post dredge-up or in the RC, and there are few stars with low $\log(g)$, the predicted evolutionary abundance corrections are insensitive to $\log(g)$ at fixed stellar mass, so these differences do not affect our analysis except to the extent that they affect the APOGEE estimates of surface abundances. 

In Fig. \ref{fig:kiel-diagram}, we also show the stellar evolutionary tracks we computed with MESA \citep{paxton2011,paxton2013,paxton2015,paxton2018}. Models with $[\alpha/\text{Fe}]=0$ are adopted in the [Fe/H]-bins centered at $\text{[Fe/H]}=0.0$ and $0.25\,\text{dex}$, whereas models with $[\alpha/\text{Fe}]=0.2$ are shown in the bin centered at $[\text{Fe/H}]=-0.25$, to reflect the observed decrease of $[\alpha/\text{Fe}]$ as the metallicity increases. We note that the models are plotted in order of decreasing stellar mass and lower mass tracks can cover over the tracks of more massive stars. 
At fixed [Fe/H], increasing $[\alpha/\text{Fe}]$ moves the models to lower effective temperatures, improving the agreement with the observations at low [Fe/H]. Fig. \ref{fig:kiel-diagram} shows that our models can qualitatively explain the observed distribution of stars in the Kiel diagram. 

The main uncertainty affecting the evolutionary tracks in Fig. \ref{fig:kiel-diagram} resides in the predicted effective temperatures of the models, which are known to be affected by systematic uncertainty, mostly due to the assumed atmosphere boundary condition and treatment of super-adiabatic convection \citep{montalban2001,salaris2002,montalban2004}. For this reason, in our analysis, the surface abundance variations of C, N, and O are characterized by considering the surface gravity and not the effective temperature. In summary, the basic stellar parameters in our analysis, which we use to characterize the properties of each star in the sample from a theoretical point of view, are the $\log(g)$ value, mass, [Fe/H] abundance, and $[\alpha/\text{Fe}]$ ratio of the stars. 

In Fig.~\ref{fig:deltaCNO_models}, we show the predictions of our MESA stellar evolution models for the evolution of the surface abundances of C, N, and O in red giants. The surface abundance change of an element $X$, $\Delta \log(X/\text{H})$, between the present-day time $t$ and the birth-time $t_{\text{birth}}$, is defined as follows:
\begin{equation}
    \Delta\log\Big(\frac{X}{\text{H}}\Big) \equiv \log \Big( \frac{N_{\text{X}}(t)}{N_{\text{H}}(t)} \Big) - \log\Big( \frac{N_{X}(t_{\text{birth}})} {N_{\text{H}}(t_{\text{birth}})} \Big),
\end{equation}
\noindent where $N_{X}$ and $N_{\text{H}}$ are the number density of atoms of the $X$ species and hydrogen, respectively. 

The top panels of Fig.~\ref{fig:deltaCNO_models} show how the surface abundances of C, N, O, and C+N are predicted to vary in red giants of different mass by our reference stellar models, for a fixed value of the surface gravity  $\log(g/[\text{cm}\,\text{s}^{-2}])=2.4$. Different curves show the predictions of models assuming different [Fe/H] abundances. We fix birth abundance ratios to solar values as we change $[\text{Fe/H}]$. In our analysis, we adopt a single value of $\log(g)$, because the surface gravity is predicted to have little effect on the surface abundance variations of C, N, O, and C+N among the red giants in our MESA models. This is demonstrated in the bottom panels of Fig.~\ref{fig:deltaCNO_models}, which show how $\Delta \log(X/\text{H})$ for $X=$ C, N, O, and C+N change as a function of $\log(g)$ for three different stellar masses ($m=1$, $1.2$, and $1.5\,\text{M}_{\sun}$), by assuming $\text{[Fe/H]}=0$. Similar results are obtained for other stellar masses and metallicities. 

The largest changes of the surface abundances are predicted for C and N, whereas the change of O is relatively small, being in the range $-0.02\lesssim \Delta \log(\text{O/H}) \lesssim 0.01$. There is a strong dependence of $\log(\text{C/H})$ and $ \log(\text{N/H})$ with stellar mass, such that C is increasingly depleted at the stellar surface as we consider red giants with increasing stellar mass, while N is increasingly enhanced, in agreement with the predictions and results of previous works (e.g., \citealt{iben1964,casali2019,shetrone2019}). For C+N, we find that $\Delta\log( (\text{C}+\text{N})/\text{H} )$ is relatively small among the red giants, being in the range $-0.05\lesssim \Delta \log(\text{(C+N)/H}) \lesssim 0.075$, and $0\lesssim \Delta \log(\text{(C+N)/H}) \lesssim 0.01$ for $m=1$-$2\,\text{M}_{\sun}$. The near-zero correction arises because extra $^{14}$N nuclei are predicted from $^{12}$C during CNO processing. 

\begin{figure}    
\centering
\includegraphics[width=7cm]{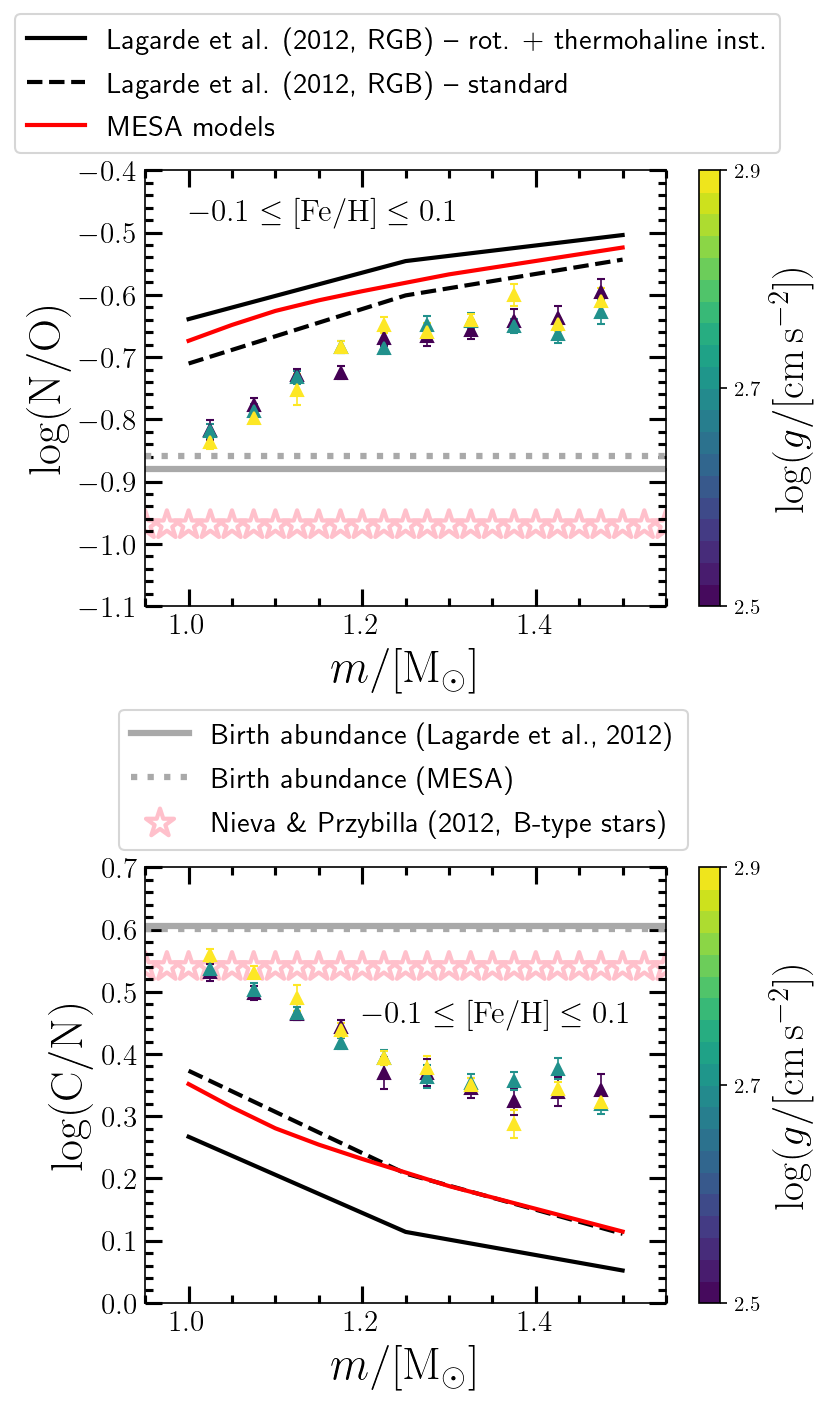}
\caption{Comparison between the predicted and observed trends of $\log(\text{N/O})$ (top panel) and $\log(\text{C/N})$ (bottom panel) as a function of the stellar mass. The black solid curves correspond to the surface abundances in RGB stars as predicted by the models of \citet{lagarde2012}, assuming rotation-induced mixing and thermohaline instability and metallicity $Z=Z_{\sun}$, whereas the the black dashed curves show the models of \citet{lagarde2012} with standard prescriptions for mixing and $Z=Z_{\sun}$. The red curve shows the prediction of our MESA stellar models for RGB stars assuming $[\text{Fe/H}]=0$. The points with error bars show the observed average trends with corresponding uncertainty on the mean that we measure in the stars classified as RGB by \citet{miglio2021}, by selecting stars with $-0.1 \le [\text{Fe/H}] \le 0.1$ and different bins of $\log(g/[\text{cm}\,\text{s}^{-2}])$ with width of $0.2$ centered at the values reported in the color bar. The models assume $\log(g/[\text{cm}\,\text{s}^{-2}])=2.6$. The gray solid line corresponds to the assumed birth abundances in the stellar models of \citet{lagarde2012}, whereas the the gray dotted line shows the assumed birth abundances in our reference MESA models. The pink star symbols show the average abundances in a sample of B-type stars in the Solar neighbourhood as measured by \citet{nieva2012}. }
\label{fig:comparison_lagarde}
\end{figure}

In Fig.~\ref{fig:comparison_lagarde}, we show how $\log(\text{N/O})$ (top panel) and $\log(\text{C/N})$ (bottom panel) change as a function of stellar mass by considering stars classified as RGB by \citet{miglio2021} with iron abundances in the range $-0.1 \le \text{[Fe/H]}\le 0.1$. The slope of the observed $\log(\text{N/O})$ and $\log(\text{C/N})$ as a function of stellar mass changes at $\approx1.2\,\text{M}_{\sun}$, which approximately divides stars with a radiative core in the MS from stars with a convective core. In the same figure, we show the predictions of the stellar models of \citet{lagarde2012} assuming $Z=Z_{\sun}$ and including either rotation-induced mixing plus thermohaline instability (black solid lines) or standard mixing prescriptions similar to our reference models (black dashed lines). We also show the predictions of our reference MESA models (red solid curves), which do not include rotation-induced mixing and thermohaline instability. The birth values of $\log(\text{N/O})$ and $\log(\text{C/N})$ assumed in the models are shown as horizontal gray lines. 

In the range of surface gravities spanned by the red giants in our sample, the predicted variation of $\log(\text{N/O})$ and $\log(\text{C/N})$ at stellar fixed mass does not depend on $\log(g)$. For this reason, the model predictions in Fig.~\ref{fig:comparison_lagarde} are shown as a function of stellar mass by assuming $\log(g/[\text{cm}\,\text{s}^{-2}])=2.6$. Surface gravity also plays a minor effect in the observed trends of $\log(\text{N/O})$ and $\log(\text{C/N})$ vs stellar mass among the RGB stars in our sample. When moving from the bin centered at $\log(g/[\text{cm}\,\text{s}^{-2}])=2.5$ to that at $\log(g/[\text{cm}\,\text{s}^{-2}])=2.9$, the differences of $\log(\text{N/O})$ and $\log(\text{C/N})$ are below $\approx0.05\,\text{dex}$ for almost all mass-bins, with the exception of RGB stars in the mass-bin $1.35 \le m < 1.4\,\text{M}_{\odot}$, which show larger abundance differences. We note that the error bars in Fig. \ref{fig:comparison_lagarde} correspond to the uncertainty on the mean. 

A difference in the surface abundances between RGB and RC stars of the same mass and metallicity is predicted by stellar evolution models including rotation-induced mixing and thermohaline instability (e.g., \citealt{lagarde2012}, and the observations of \citealt{taut2010} for $^{12}$C/$^{13}$C), whereas little or no difference is predicted by stellar models with the prescriptions described in Section \ref{sec:MESA-models} 
(see the bottom panels of Fig. \ref{fig:deltaCNO_models}). For this reason, our analysis in Fig.~\ref{fig:comparison_lagarde} focuses only on RGB stars. We discuss the differences between RGB and RC stars in Section \ref{sec:RGB-vs-RC}.

In Fig.~\ref{fig:comparison_lagarde}, the trends of $\log(\text{N/O})$ and $\log(\text{C/N})$ vs stellar mass as predicted by our set of stellar models qualitatively agree  with the average trends that we see in the observational data for the RGB stars in our sample. Models with rotation-induced mixing and thermohaline instability predict larger depletion of $\log(\text{C/N})$ and enhancement of $\log(\text{N/O})$ than the models with standard mixing prescriptions. Furthermore, our reference MESA stellar evolution models predict the same $\log(\text{C/N})$ and slightly larger $\log(\text{N/O})$ by $\approx0.04\,\text{dex}$ than the models of \citet{lagarde2012} with standard mixing prescriptions, mostly due to the slightly different $\log(\text{N/O})$ assumed at birth in the two models. Quantitatively, all of the models in Fig.~\ref{fig:comparison_lagarde} predict values of $\log(\text{N/O})$ that are too high and values of $\log(\text{C/N})$ that are too low relative to APOGEE measurements. For $m=1.3\,\text{M}_{\sun}$, the discrepancies with the MESA models are $\approx0.12\,\text{dex}$ and $\approx0.08\,\text{dex}$, respectively. The differences increase towards $m=1\,\text{M}_{\sun}$. These discrepancies could indicate that the models predict too much mixing, but they could also arise from incorrect assumptions about the birth abundances of C, N, and O, or from systematic errors in the APOGEE measurements.

\begin{figure}    
\centering
\includegraphics[width=7cm]{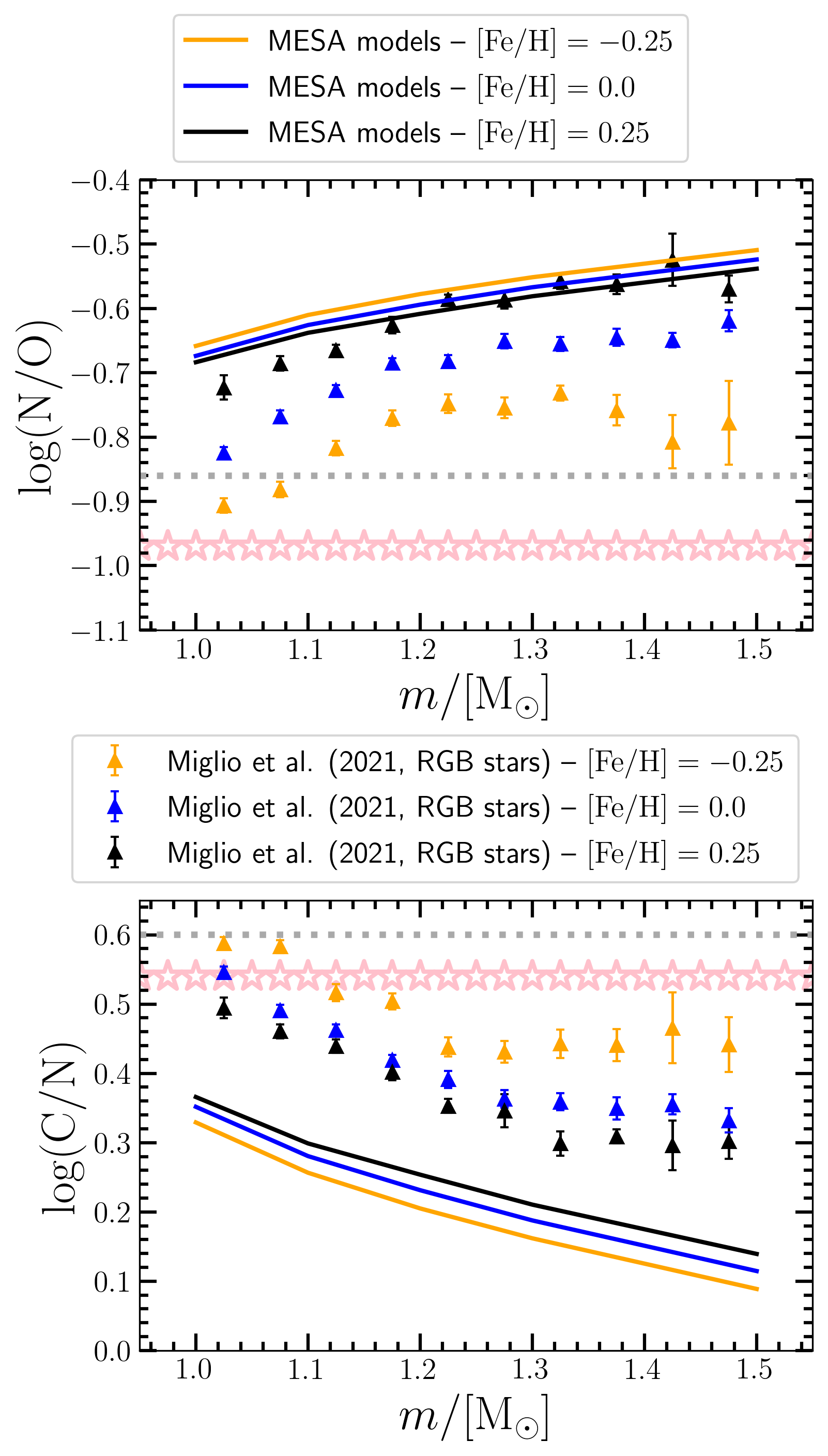}
\caption{\textit{Top panel}: observed average $\log(\text{N/O})$ with corresponding uncertainty on the mean as a function of the stellar mass in the sample of stars classified as RGB by \citet{miglio2021} (triangles with error bars), by selecting stars with $\log(g/[\text{cm}\,\text{s}^{-2}])=2.6\pm0.1$. Different colours correspond to different bins of [Fe/H] with width $\Delta [\text{Fe/H}]=0.25$, centered at the value reported in the legend. In the figure, the observations are compared with the predictions of our reference MESA stellar evolution models (solid curves with different colours). The gray horizontal line shows the assumed birth abundances in our MESA models, whereas the pink star symbols show the average abundances in a sample of B-type stars in the Solar neighbourhood as measured by \citet{nieva2012}. \textit{Bottom panel}: same as in top panel but showing our results for $\log(\text{C/N})$ vs stellar mass. The MESA models adopt $[\alpha/\text{Fe}]=0$ for all three values of [Fe/H]. 
}
\label{fig:NO-CN-logg}
\end{figure}

The assumed birth abundances are based on solar ratios. \citealt{lagarde2012} adopt birth values $\log(\text{N/O}) \approx -0.86$ and $\log(\text{C/N}) \approx 0.60$, and the MESA models have $\log(\text{N/O}) \approx -0.88$. \citet{nieva2012} find $\log(\text{N/O})\approx-0.97$ and $\log(\text{C/N})\approx0.54$ in a sample of B-type stars in the Solar neighbourhood. 
The difference from solar ratios could reflect a change with age at fixed $[\text{Fe/H}]$, or errors in the solar abundance or B-star calibrations. The differences between the assumed birth abundances and the measurements of \citet{nieva2012} are similar in magnitude to the discrepancies between the standard mixing models and the data in Fig.~\ref{fig:comparison_lagarde}, and they have the correct sign to explain the $\log(\text{N/O})$ discrepancy but not the $\log(\text{C/N})$ discrepancy. The impact of changing birth abundances is not simply additive, and in Section \ref{sec:different-birth-CNO} below we use an approximate method to assess whether a reasonable change in birth abundances can reconcile the models with the APOGEE data. The \citet{lagarde2012} models with extra mixing predict larger $\log(\text{N/O})$ enhancement and $\log(\text{C/N})$ depletion, producing greater tension with the APOGEE data.

From a nucleosynthesis point of view, $\log(\text{C/N})$ and $\log(\text{N/O})$ are expected to have a dependence with metallicity, because of the strong metallicity-dependence of the N nucleosynthesis in AGB and massive stars (e.g., see \citealt{vincenzo2016,vincenzo2018a,vincenzo2018b}). In particular, $^{14}\text{N}$ can be synthesized in the CNO cycle at the expense of the C and O nuclei present in the star at its birth. For this reason, in the late evolutionary phases of galaxies, where the chemical enrichment of AGB stars dominates N production, $\log(\text{N/O})$ is predicted to steadily increase in the gas-phase, whereas $\log(\text{C/N})$ decreases. 

Metallicity has an effect on the predicted abundance variations of C, N, and O in red giants. This is illustrated in Fig.~\ref{fig:NO-CN-logg}, which shows how $\log(\text{N/O})$ (top panel) and $\log(\text{C/N})$ (bottom panel) change as a function of stellar mass in our sample, when considering different [Fe/H]-bins. At fixed stellar mass, stars with higher [Fe/H] are observed to have larger $\log(\text{N/O})$ and lower $\log(\text{C/N})$. The surface abundance variations of $\log(\text{N/O})$ and $\log(\text{C/N})$ as predicted by our reference MESA models show an opposite trend as a function of metallicity with respect to observations, but the predicted metallicity-dependence of the atmospheric $\log(\text{N/O})$ and $\log(\text{C/N})$ abundance ratios in the models is much weaker than the dependence that we see in the observations. We note that the birth abundance ratios of $\log(\text{C/N})$ and $\log(\text{N/O})$ assumed in the models are constant as a function of [Fe/H], because the models assume scaled solar composition. 
Since the birth abundance ratios of $\log(\text{N/O})$ and $\log(\text{C/N})$ are metallicity-independent, the opposite trend of the predicted surface abundance ratio changes with increasing metallicity in Fig.~\ref{fig:NO-CN-logg} implies that metal-rich stars are (moderately) less affected by mixing processes than metal-poor stars.  

The effects of metallicity in the models of \citet{lagarde2012} with rotation-induced mixing and thermohaline instability are consistent with the predictions of our MESA models (Fig. \ref{fig:NO-CN-logg}). By assuming the stellar models of \citet{lagarde2012}, the variations due to metallicity are in the range  $0.03\lesssim|\langle \Delta\log(\text{N/O})_{\text{mod}}\rangle|\lesssim0.06$ and $0.05\lesssim|\langle \Delta\log(\text{C/N})_{\text{mod}}\rangle|\lesssim 0.12$ when moving from $Z=0.004$ to  $Z=0.014$, and $0.04\lesssim|\langle \Delta\log(\text{N/O})_{\text{mod}}\rangle|\lesssim0.1$ and $0.07\lesssim|\langle \Delta\log(\text{C/N})_{\text{mod}}\rangle|\lesssim 0.17$ between $Z=0.002$ and $Z=0.014$. From these values, the predicted dependence of $\log(\text{C/N})$ with metallicity is almost two times larger than that of $\log(\text{N/O})$. 

The stellar models of \citet{lagarde2012} assume that the rotation velocity does not vary as a function of metallicity, being in the range between $90$ and $137\,\text{km}\,\text{s}^{-1}$ on the main sequence.  Recently, \citet{amard2020} found that, at fixed stellar mass, the average rotation velocity decreases as a function of metallicity in a sample of stars in \textit{Kepler} (see fig. 6 of \citealt{amard2020}), in addition to the dependence of the average rotation period with stellar mass. \citet{charbonnel2010} showed that models with higher rotation velocities at solar metallicity predict a larger depletion of $\log(\text{C/N})$. Therefore, if we extrapolate the findings of \citet{charbonnel2010} for C/N vs $v_{\text{rot}}$ at low metallicity, the effect of rotation would amplify the opposite  dependence on metallicity which is seen between models and observations in Fig. \ref{fig:NO-CN-logg}. This corroborates our conclusion that the trend that we see in the observations is dominated by chemical evolution of birth abundances and not by mixing processes inside the stars.

\begin{figure}    
\centering
\includegraphics[width=7cm]{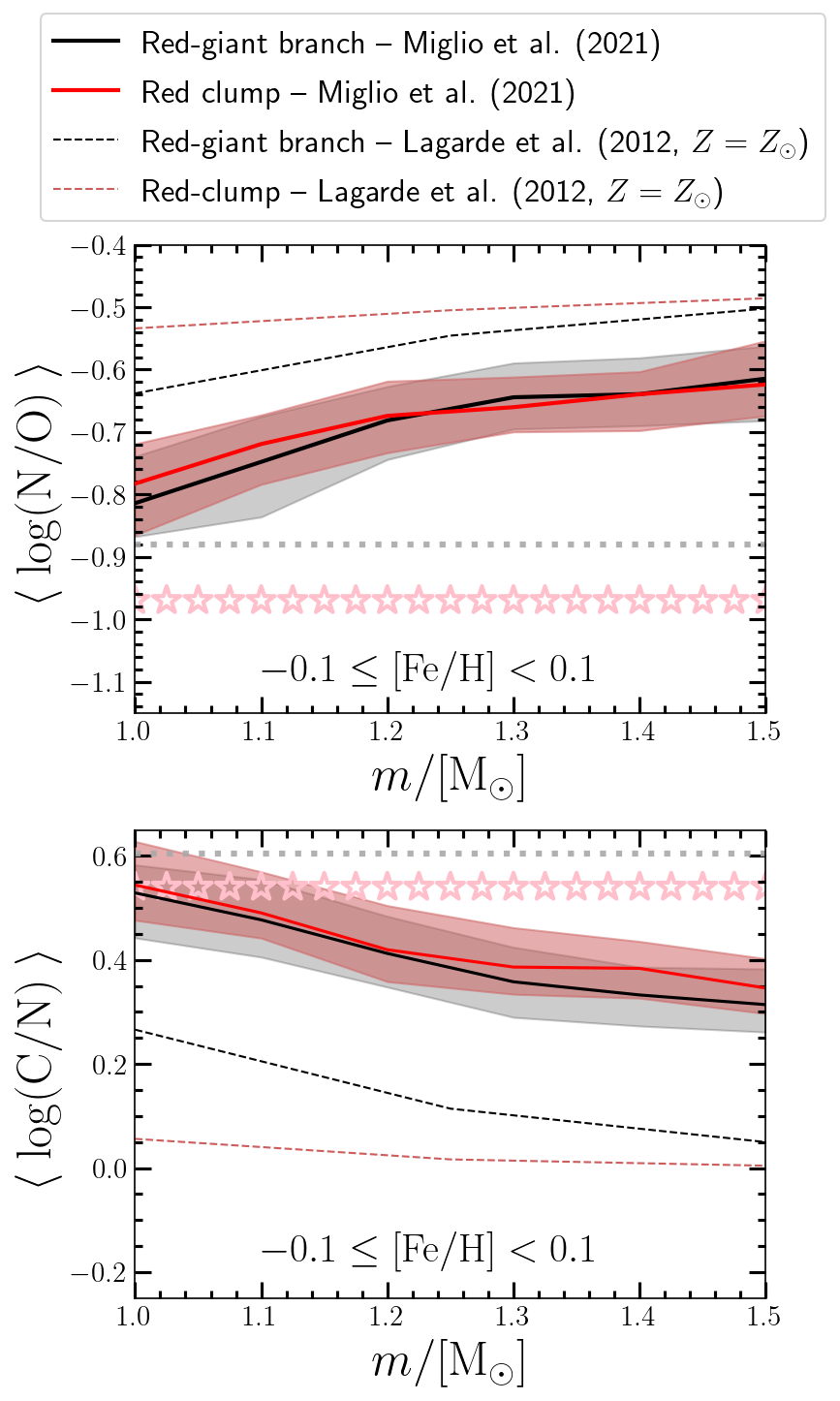}
\caption{The observed median trends of $\log(\text{N/O})$ (top panel) and $\log(\text{C/N})$ (bottom panel) with corresponding $16$ and $84$ percentiles as a function of the stellar mass as computed from the abundance distributions of stars classified as RGB (black curves with shaded areas) and RC (red curves with shaded areas) by \citet{miglio2021}, in the range of iron abundances $-0.1 \le [\text{Fe/H}] \le 0.1$. The black and red dotted curves correspond to the surface abundance ratios as predicted by the stellar models of \citet{lagarde2012} with rotation-induced mixing and thermohaline instability for RGB and RC stars, respectively, at $Z=Z_{\sun}$. The pink star symbols show the average abundances in a sample of B-type stars in the Solar neighbourhood as measured by \citet{nieva2012}.   }
\label{fig:variation-rc-vs-rgb}
\end{figure}

\subsection{Surface abundance distributions of red-giant branch vs red clump stars}
\label{sec:RGB-vs-RC}

Most of the stars in the \citet{miglio2021} sample also have information available about their evolutionary state (for example, if they are in the RGB or RC), which was determined using the approach defined in \citet{elsworth2019}. In our analysis, we select stars in \citet{miglio2021} with [Fe/H] in the range $-0.1\le\text{[Fe/H]}<0.1$ and mass in the range $0.9\le m <1.6\,\text{M}_{\sun}$, and divide them in mass-bins $\Delta m =0.1\,\text{M}_{\sun}$. Then, for each mass-bin, we compute the distribution of $\log(\text{N/O})$ and $\log(\text{C/N})$ of RGB and RC stars separately, using the abundance measurements from APOGEE-DR16 and the classification provided by \citet{miglio2021}. For each distribution, we compute the median values -- together with the $16$ and $84$ percentiles -- of the distributions of $\log(\text{N/O})$ and $\log(\text{C/N})$ for RGB and RC stars, separately. 

The results of our analysis are illustrated in Fig. \ref{fig:variation-rc-vs-rgb}, in which the median values of $\log(\text{N/O})$ (top panel) and $\log(\text{C/N})$ (bottom panel) in RGB stars are compared with the corresponding values derived in RC stars. The shaded areas in the figures correspond to the $16$ and $84$ percentiles of the distributions around the median values. In the same figure, we also show the predictions of the stellar models of \citet{lagarde2012} assuming $Z=Z_{\sun}$ and rotation-induced mixing plus thermohaline instability. The predictions of the models for RGB stars are shown as black dotted lines, whereas the predictions for RC stars are the red dotted lines. Similar results are found for different [Fe/H]-bins, but \citet{lagarde2012} provide stellar models over a relatively coarse grid in mass and metallicity. We do not show the predictions of our MESA stellar models, because they do not predict any difference in $\log(\text{N/O})$ and $\log(\text{C/N})$ between RGB and RC stars (see the bottom panels in Fig. \ref{fig:deltaCNO_models}). 

The results of our analysis in Fig. \ref{fig:variation-rc-vs-rgb} show that RGB and RC stars have similar $\log(\text{N/O})$ and $\log(\text{C/N})$ as a function of stellar mass. This is in disagreement with the predictions of the models of \citet{lagarde2012} with extra-mixing, which predict RC stars to have systematically higher $\log(\text{N/O})$ and lower $\log(\text{C/N})$ than RGB stars at all stellar masses. We note that, at high stellar masses, the observed $\log(\text{C/N})$ in RC stars are slightly larger than in RGB stars, which is the opposite of what the models of \citet{lagarde2012} predict.

\subsection{Making Model Predictions with Different Birth CNO}
\label{sec:different-birth-CNO}

The stellar models predict $\log(\text{C/H})$, $\log(\text{N/H})$, and $\log(\text{O/H})$, given the mass, metallicity, $[\alpha/\text{Fe}]$, and birth abundances of the stars. However, the birth abundances of young solar metallicity stars (e.g., \citealt{nieva2012}) may be different from the Sun's photospheric abundances (e.g., \citealt{asplund2009}). Uncertainty in the CNO birth abundances of the order of $\approx 0.1\,\text{dex}$ can affect the success of stellar models in reproducing the observed abundances of C, N, and O in red giants, as shown in Fig. \ref{fig:NO-CN-logg}. 

In this Section, we provide a formalism to estimate the values of $\log(\text{C/H})$, $\log(\text{N/H})$, and $\log(\text{O/H})$ that one would get with different birth abundances, by knowing the results for the solar birth abundances used in the MESA models. 
Our approximation is based on the assumptions that \textit{(i)} the material mixed into the envelope is independent of birth abundances and \textit{(ii)} the impact of birth abundances is only to change what the dredged-up material is being diluted by. 

We adopt the following notation for the assumed birth abundance of the elements $\mathcal{Z}=$ C, N, and O in the model:
\begin{equation} \label{eq3}
    \mathcal{Z}_{\text{b},0} \equiv  (\mathcal{Z}/\text{H})_{\text{b},0}, 
\end{equation}
in which H is dropped for notational convenience. The baseline model predictions can be expressed as follows: 
\begin{equation} \label{eq4}
    \mathcal{Z}_{0} = \mathcal{Z}_{\text{b},0} + \Delta \mathcal{Z},
\end{equation}
where $\Delta \mathcal{Z}$ represents the change in the abundance of  $\mathcal{Z}$ per unit hydrogen as predicted by MESA models because processed material is mixed into the envelope. We note that $\Delta \mathcal{Z}$ is not in dex.  

We allow a difference $\delta \mathcal{Z}_{\text{b}}$ in the birth abundances relative to the model assumptions. We specify these differences in dex, so that the new birth abundances are determined by the following expression:
\begin{equation} \label{eq5}
\mathcal{Z}_{\text{b}} = \mathcal{Z}_{\text{b},0} \times 10^{\delta \mathcal{Z}_{\text{b}}}.
\end{equation}
We want to know if we can reconcile the models with APOGEE data in Fig. \ref{fig:NO-CN-logg}, by allowing reasonable choices for $\delta \mathcal{Z}_{\text{b}}$ while assuming that the models correctly predict $\Delta \mathcal{Z}$. 

The original model predictions are 
\begin{equation} \label{eq6}
    \log(\mathcal{Z}/\text{H}) = \log( \mathcal{Z}_{\text{b},0} + \Delta \mathcal{Z} ). 
\end{equation} 
With a perturbation $\delta \mathcal{Z}_{\text{b}}$ in the birth abundances (in dex), the prediction becomes
\begin{equation} \label{eq7}
\begin{aligned}
\mathcal{Z} & = \mathcal{Z}_{\text{b}} \times 10^{ \delta \mathcal{Z}_{\text{b}} } + \Delta \mathcal{Z} \\
 & = \mathcal{Z}_{\text{b},0} + \Delta \mathcal{Z} + \mathcal{Z}_{\text{b},0} \times \Big( 10^{ \delta \mathcal{Z}_{\text{b}} } - 1 \Big), 
\end{aligned}
\end{equation} 
in which we have used equation \ref{eq5}. 

By putting together equations \ref{eq4} and \ref{eq7}, we find the following equation for the change of $\mathcal{Z}$ after a perturbation $\delta \mathcal{Z}_{\text{b}}$ is applied to the assumed birth abundances: 
\begin{equation} \label{eq8}
\begin{aligned}
\log( \mathcal{Z}/\text{H} ) & - \log( \mathcal{Z}/\text{H} )_{0} = \\ 
& = \log\Big[ \frac{ \mathcal{Z}_{\text{b},0} + \Delta \mathcal{Z} + \mathcal{Z}_{\text{b},0} \times ( 10^{ \delta \mathcal{Z}_{\text{b}} } - 1 ) }{ \mathcal{Z}_{\text{b},0} + \Delta \mathcal{Z} } \Big]  \\
& = \log\Big[ 1 + 10^{ -\Delta\log( \mathcal{Z}/\text{H} )_{0} } ( 10^{ \delta \mathcal{Z}_{\text{b}} } - 1 ) \Big], 
\end{aligned}
\end{equation} 
where 
\begin{equation} \label{eq9}
    \begin{aligned}
    \Delta \log( \mathcal{Z}/\text{H} )_{0} & = \log( \mathcal{Z}/\text{H} )_{0} - \log( \mathcal{Z}/\text{H} )_{\text{b},0} \\
    & = \log\Big[  1 + \frac{ \Delta \mathcal{Z} }{ \mathcal{Z}_{\text{b},0}  }  \Big]. 
    \end{aligned}
\end{equation} Qualitatively, equation \ref{eq8} explains why, in Fig. \ref{fig:NO-CN-logg}, the discrepancy is larger for small masses than for large masses. Because $\Delta\log(\text{N/H})_{0}$ increases with mass in Fig. \ref{fig:deltaCNO_models}, the correction expressed by equation \ref{eq8} is bigger when the mass is smaller.

\begin{figure}    
\centering
\includegraphics[width=7cm]{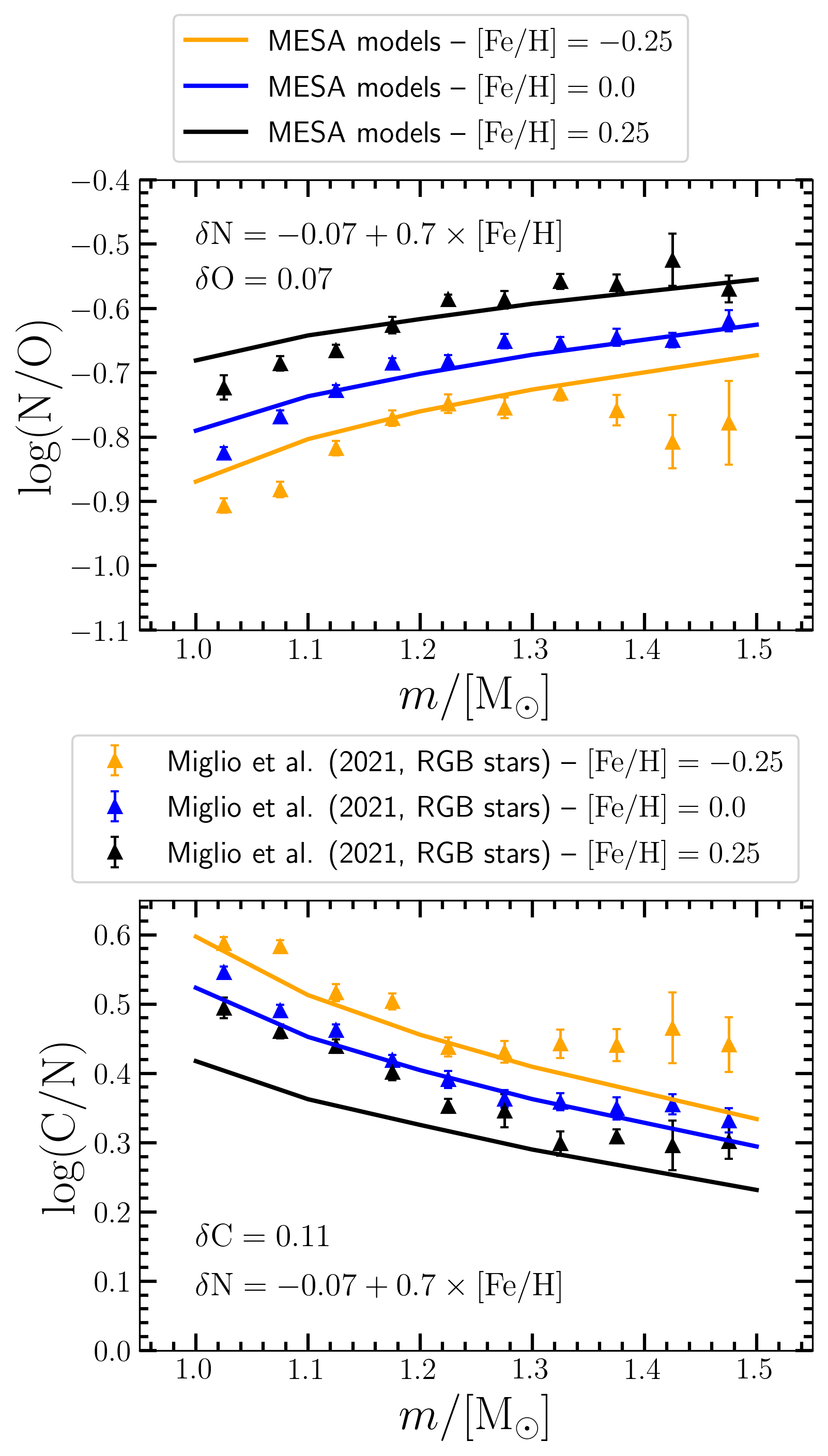}
\caption{
Similar to Fig. \ref{fig:NO-CN-logg} but showing the effect of perturbing the assumed birth abundances at $Z=Z_{\sun}$ in our MESA models according to the formalism outlined in Section \ref{sec:different-birth-CNO}, in order to reconcile the observed relations of $\log(\text{N/O})$ vs mass (top panel) and $\log(\text{C/N})$ vs mass (bottom panel) with the models predictions. 
}
\label{fig:CN-NO-corrections}
\end{figure}

\begin{figure*}    
\centering
\includegraphics[width=12.5cm]{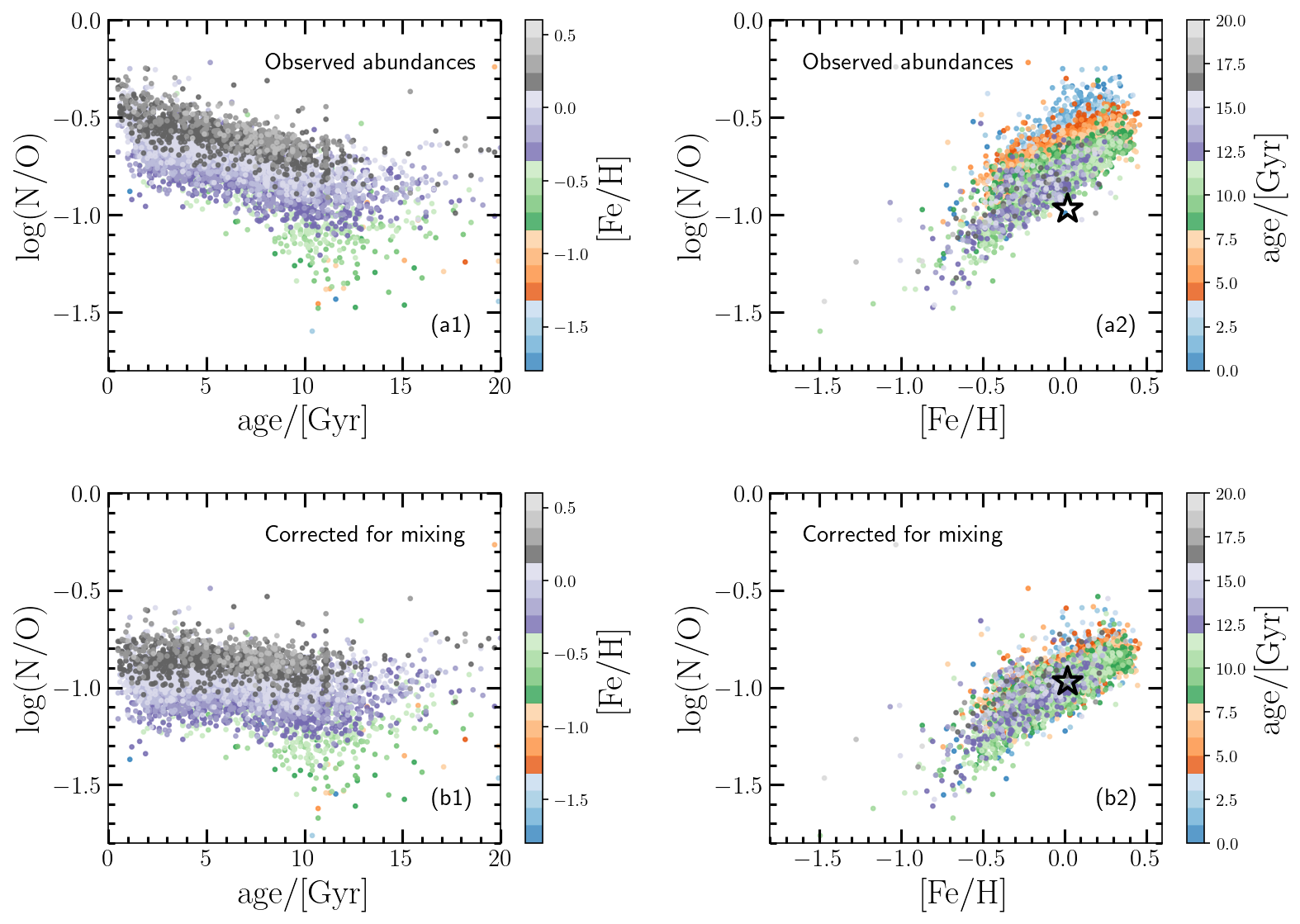}
\caption{\textit{Panel (a1)}: observed $\log(\text{N/O})$ vs age, by using the ages from the analysis of \citet{miglio2021} and the abundances from APOGEE-DR16; the color-coding corresponds to the [Fe/H] abundance of the stars. \textit{Panel (a2)}: observed $\log(\text{N/O})$ vs [Fe/H], with the color-coding representing the ages of the stars from the analysis of \citet{miglio2021}. The black star shows the observed average abundances in a sample of B-type stars in the Solar neighbourhood by \citet{nieva2012}. \textit{Panel (b1)}: predicted $\log(\text{N/O})$ vs age when the surface abundances of C and N are corrected for mixing effects using the predictions of our reference MESA stellar evolution models. \textit{Panel (b2)}: predicted $\log(\text{N/O})$ vs [Fe/H] when the surface abundances of C and N are corrected for mixing effects as in panel (b1). The black star shows the results of the abundance analysis of \citet{nieva2012} for a homogeneous sample of B-type stars in the Solar neighbourhood.}
\label{fig:NO-plots}
\end{figure*}

\begin{figure*}    
\centering
\includegraphics[width=12.5cm]{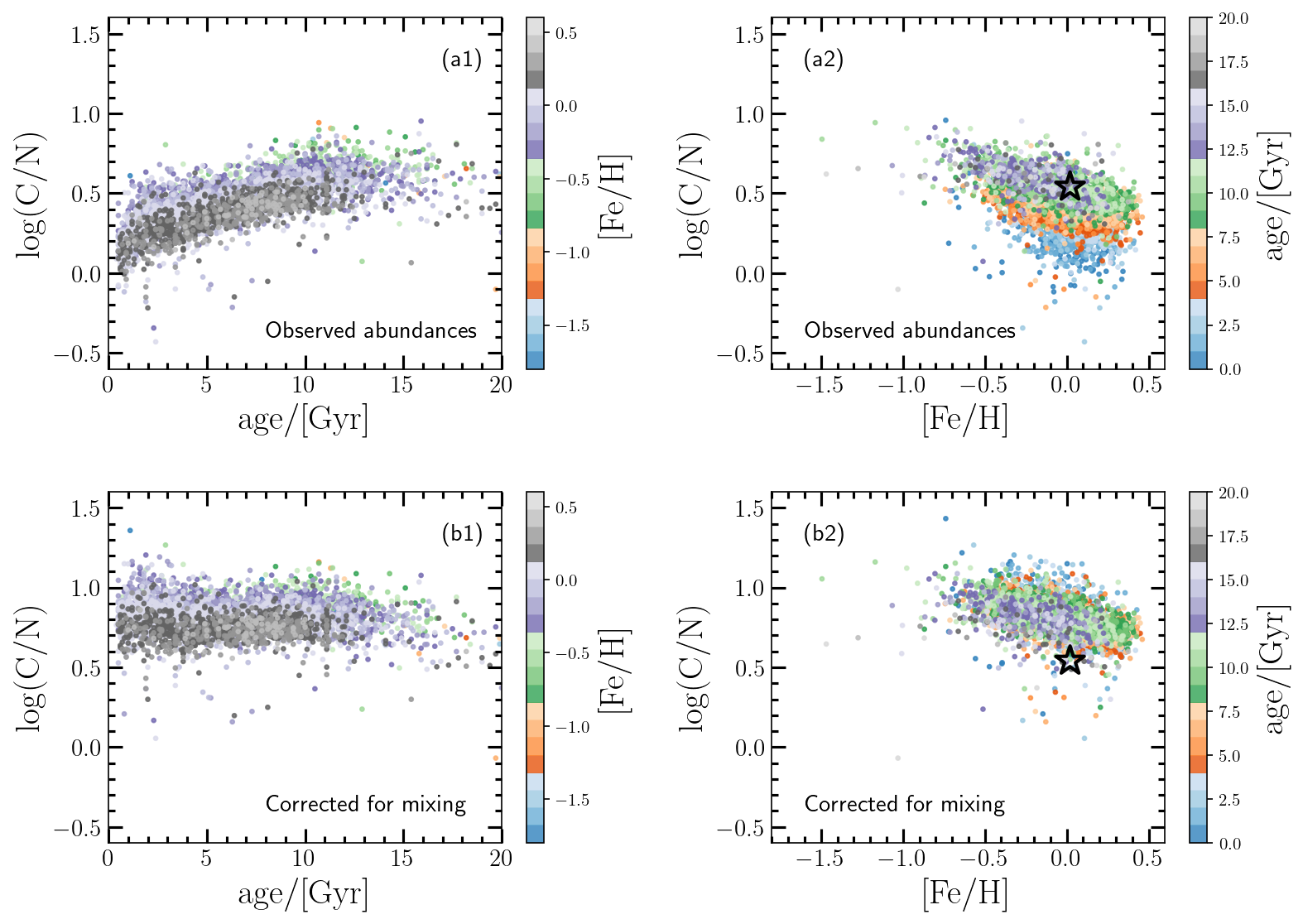}
\caption{Similar set of panels as in Fig. \ref{fig:NO-plots} but showing the results of our analysis for the $\log(\text{C/N})$ vs age and [Fe/H] relations in the sample of \citet{miglio2021}. The black star in (a2) and (b2) shows the observed average abundances in a sample of B-type stars in the Solar neighbourhood by \citet{nieva2012}.}
\label{fig:CN-plots}
\end{figure*}

In our analysis, we assume that N birth abundances scale with metallicity according to the following relation:
\begin{equation} \label{eq10}
    \delta \text{N}_{\text{b}} =  \delta \text{N}_{\sun} + 0.7 \times [\text{Fe/H}],
\end{equation}
where $\delta\text{N}_{\sun}$ corresponds to the perturbation on the N birth abundance at $Z=Z_{\sun}$ assumed by our MESA models taken from \citet{asplund2009}, and the slope of $0.7$ is based on our findings in Fig. \ref{fig:comparison-gas-phase} (see also Section \ref{sec:comparing} for a detailed discussion), which shows abundance measurements of $\log(\text{N/O})$ in the ISM of galaxies and the calibration relation for the gas-phase $\log(\text{N/O})$ vs $\log(\text{O/H})+12$ proposed by \citet{dopita2016}.

For C and O, we assume that the birth abundances do not vary with metallicity:  
\begin{equation}
\begin{aligned}
\delta \text{C}_{\text{b}} & =  \delta \text{C}_{\sun}, \\ 
\delta \text{O}_{\text{b}} & =  \delta \text{O}_{\sun}.
\end{aligned}
\end{equation}

In Fig. \ref{fig:CN-NO-corrections}, we show that MESA predictions can be reconciled with the observed $\log(\text{N/O})$ vs mass and $\log(\text{C/N})$ vs mass in different metallicity ranges by assuming $\delta\text{C}_{\sun}=0.11$, $\delta\text{N}_{\sun}=-0.07$, and $\delta\text{O}_{\sun}=0.07$. These values of $\delta\text{N}_{\sun}$ and $\delta\text{O}_{\sun}$ relative to \citet{asplund2009} are similar to the differences between the abundances of \citet{nieva2012} (who measured $\log(\text{C/H})=8.33$, $\log(\text{N/H})=7.79$, $\log\text{(O/H)}=8.76$ in nearby B-type stars) and those of \citet{asplund2009} (who measured $\log(\text{C/H})=8.43$, $\log(\text{N/H})=7.83$, $\log\text{(O/H)}=8.69$ in the Sun photosphere). However, they differ for $\delta\text{C}_{\sun}$ by $0.2\,\text{dex}$, since the difference between \citet{nieva2012} and \citet{asplund2009} is $-0.1\,\text{dex}$ for C.  

Fig. \ref{fig:CN-NO-corrections} shows agreement with the APOGEE $\log(\text{N/O})$ and $\log(\text{C/N})$ at the $\approx0.05\,\text{dex}$ level for most masses and metallicities, with differences rising to $\approx0.1\,\text{dex}$ in some bins. This agreement shows that the larger differences seen in Figs. \ref{fig:comparison_lagarde} and  \ref{fig:NO-CN-logg} are plausibly explained by incorrect assumptions about birth abundances rather than incorrect mixing predictions, perhaps in combination with zero-point calibration offsets in the abundance measurements themselves. However, it does not guarantee that the mixing predictions are accurate. The success of this assumption in explaining the observed metallicity dependence of $\log(\text{N/O})$ and $\log(\text{C/N})$ supports our conjecture that chemical evolution of N birth abundances dominates these trends.

\section{Correcting APOGEE abundances for evolutionary effects}
\label{sec:correcting}

The predictions of our reference MESA stellar evolution models are used to compute the surface abundance changes $\Delta \log(\text{C/H})$, $\Delta \log(\text{N/H})$, and $\Delta \log(\text{O/H})$ in the stars of our sample, depending on their surface gravity, mass, [Fe/H] abundance, and $[\alpha/\text{Fe}]$ ratio. We do not interpolate or extrapolate in the grid of stellar evolution tracks, but use the closest point to compute a theoretical correction due to mixing for each star in the sample. The $[\alpha/\text{Fe}]$ ratio of the stars is taken from the quantity 
\texttt{ALPHA\_M} in the APOGEE-DR16 catalogue. 

If our abundance and mass measurements and our mixing models were perfectly accurate, it would be preferable to correct $\Delta\text{N}$, $\Delta\text{C}$, and $\Delta\text{O}$ as described in Section \ref{sec:different-birth-CNO}. However, if the mixing corrections is comparable to the observed surface abundance, then the implied change in $\log(\text{C/H})$, $\log(\text{N/H})$, and $\log(\text{O/H})$ can become very large, and the implied birth abundance can even become negative. We have therefore chosen to apply the $\Delta\log(\text{C/H})$, $\Delta\log(\text{N/H})$, and $\Delta\log(\text{O/H})$ predicted by the models, which we find yields similar results on average and is more robust to imperfections in the data and models. 

\subsection{The variation of $\log(\text{N/O})$}

The results of our analysis for $\log(\text{N/O})$ are illustrated in Fig. \ref{fig:NO-plots}, which compares $\log(\text{N/O})$ vs age (panel a1) and $\log(\text{N/O})$ vs $[\text{Fe/H}]$ (panel a2) as observed in the stars of our sample with the trends corrected for mixing effects following our reference MESA stellar evolution models (panels b1 and b2). 

Without a correction, a strong correlation is observed between $\log(\text{N/O})$ and age at fixed [Fe/H] abundance (see panels a1 and a2). When we apply the abundance corrections due to mixing as predicted by our reference MESA stellar models, the age-dependence of $\log(\text{N/O})$ is effectively removed for ages $\lesssim10\,\text{Gyr}$ (see panels b1 and b2). At older ages, there is a population of metal-poor red giants with corrected $\log(\text{N/O})\approx -1.5$, consistent with the observations in metal-poor halo stars (e.g., \citealt{spite2005}), damped-Ly$\alpha$ systems (e.g. \citealt{pettini2002,pettini2008}), and metal-poor star-forming blue compact dwarf galaxies (e.g., \citealt{berg2012,izotov2012,james2015}). The plateau in $\log(\text{N/O})$ at low metallicity has been used to infer the presence of a primary N component from massive stars (see \citealt{matteucci1986,chiappini2003,chiappini2005,vincenzo2016}). 

After correcting the observed $\log(\text{N/O})$ for the mixing, we are left with a strong dependence of $\log(\text{N/O})$ on $\text{[Fe/H]}$. This is in agreement with observations in nearby spiral galaxies, which show striking relations between $\log(\text{N/O})$ and $\log(\text{O/H})+12$ (e.g. \citealt{henry2000,pettini2002,pettini2008,pilyugin2010,belfiore2017,berg2020}), and stellar abundance measurements in our Galaxy (e.g., \citealt{magrini2018} with Gaia-ESO survey). As discussed by \citet{vincenzo2018b}, the $\log(\text{N/O})$ vs. metallicity diagram is regulated -- to first order -- by the nucleosynthesis of N, which strongly depends on metallicity, being mainly synthesized in the CNO cycle at the expense of the C and O nuclei already present in the stars at birth. Secondary effects regulating the evolution of stellar and gas-phase abundances in $\log(\text{N/O})$ vs $\log(\text{O/H})+12$ can be variations in the star formation efficiency and bursts of star formation caused by sudden inflows of gas, because of the different time scales with which N and O are released by dying stellar populations in galaxies (see \citealt{vincenzo2016}). 

In Fig. \ref{fig:NO-plots}, we show that our predictions for the corrected $\log(\text{N/O})$ vs $\text{[Fe/H]}$ are consistent with the average measurements of \citet{nieva2012} in a sample of B-type stars in the Solar neighbourhood, providing indication that our reference MESA models are able to effectively capture the main mixing evolutionary processes affecting N abundances in stars.  

\subsection{The variation of $\log(\text{C/N})$}

The results of our analysis for $\log(\text{C/N})$ are shown in Fig. \ref{fig:CN-plots}. Similarly to $\log(\text{N/O})$, there is a strong dependence of observed $\log(\text{C/N})$ as a function of age in our sample; moreover, at fixed age, stars with increasing metallicities are observed to have higher $\log(\text{C/N})$. The strong correlation between $\log(\text{C/N})$  and age/mass in red giants is known to be due to internal mixing processes, which are more important in stars with larger masses, explaining why $\log(\text{C/N})$ is used as an age diagnostic for red giants (e.g., \citealt{iben1964,salaris2015,masseron2015,martig2016,miglio2021}). When we correct the observed $\log(\text{C/N})$ to account for the mixing processes, we effectively remove the strong age-dependence of $\log(\text{C/N})$, and only the metallicity dependence is left (see panels b1-b2), such that stars with higher metallicities have -- on average -- lower $\log(\text{C/N})$. This is expected from a nucleosynthesis point of view because of the opposite metallicity-dependence of the C and N yields from AGB and massive stars (e.g., \citealt{vincenzo2018a}). Models with diffusive convective core overshooting efficiency $f_{\text{ov,core}}=0.02$ predict that stars with mass $m\ge2\,\text{M}_{\sun}$ have their N abundances after the first dredge-up enhanced by $\approx0.04\,\text{dex}$ with respect to our reference models assuming $f_{\text{ov,core}}=0$; this could help flattening the small bump in the corrected $\text{C/N}$ ratios at ages about $\approx1\,\text{Gyr}$ and $\text{[Fe/H]}\approx 0$.

The mixing-corrected $\log(\text{C/N})$ ratios as predicted by our reference MESA models lie above the average measurements of \citet{nieva2012} at $[\text{Fe/H}]\approx 0$ in a sample of B-type stars in the Solar neighbourhood. Since the main effect of the mixing processes is enhancing the atmospheric N abundances (see Fig. \ref{fig:deltaCNO_models}) at the expense of C and O, it is difficult to reconcile the fact that our models are able to reproduce the enhancement of $\log(\text{N/O})$ but overestimate the depletion of $\log(\text{C/N})$. Systematic uncertainties of the order of $\approx0.05\,\text{dex}$ in the C and N abundance measurements of APOGEE-DR16, such that C would be increased and N decreased, would help reduce the discrepancy that we find between the mixing-corrected $\log(\text{C/N})$ as predicted by our MESA models and the average measurement of \citet{nieva2012}.

\begin{figure*}    
\centering
\includegraphics[width=12.5cm]{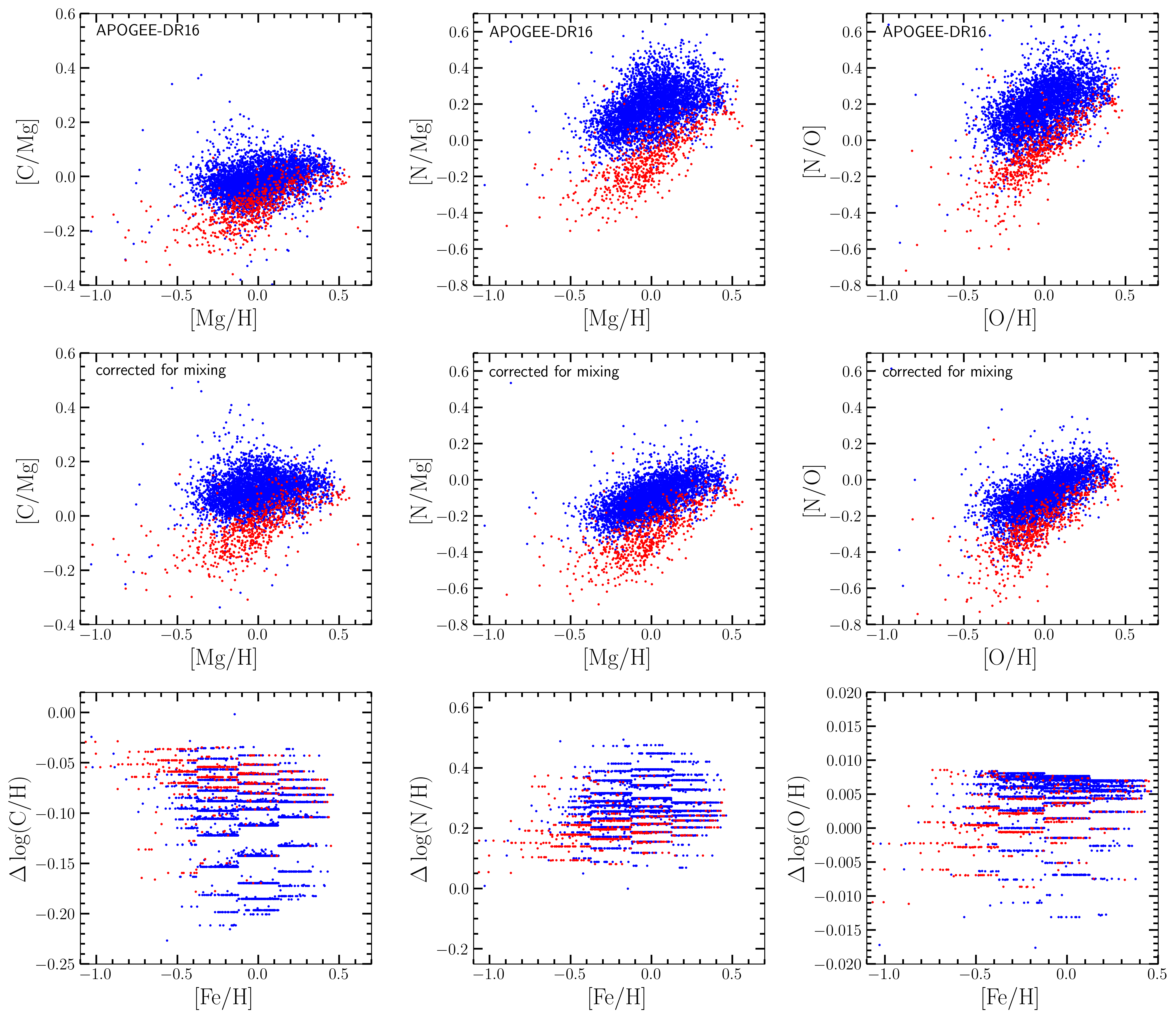}
\caption{\textit{Top panels}: the observed [C/Mg] vs [Mg/H] (first column), [N/Mg] vs [Mg/H] (second column), and [N/O] vs [O/H] (third column) as observed in APOGEE-DR16, by using the sample of stars of \citet{miglio2021}; to quantify the impact of evolutionary mixing effects in the two sequences, stars in the high-$\alpha$ component are shown in red, whereas stars in the low-$\alpha$ component are shown in blue, by using equation \ref{eq:high-low-alpha} reported in the main text.
\textit{Middle panels}: same as in the top panels, but the evolutionary effects due to mixing are taken into account by using our reference MESA stellar evolution models. Note the different vertical axis ranges for the left panels, representing the small dynamic range of [C/Mg]. \textit{Bottom panels}: $\Delta{\log(\text{C/H})}$, $\Delta{\log(\text{N/H})}$, and $\Delta{\log(\text{O/H})}$ as predicted for the stars when passing from the top to the bottom panel, by accounting for the evolutionary effects on the surface abundances due to mixing processes. 
}
\label{fig:CNO-high-low-alpha}
\end{figure*}

\begin{figure*}    
\centering
\includegraphics[width=14cm]{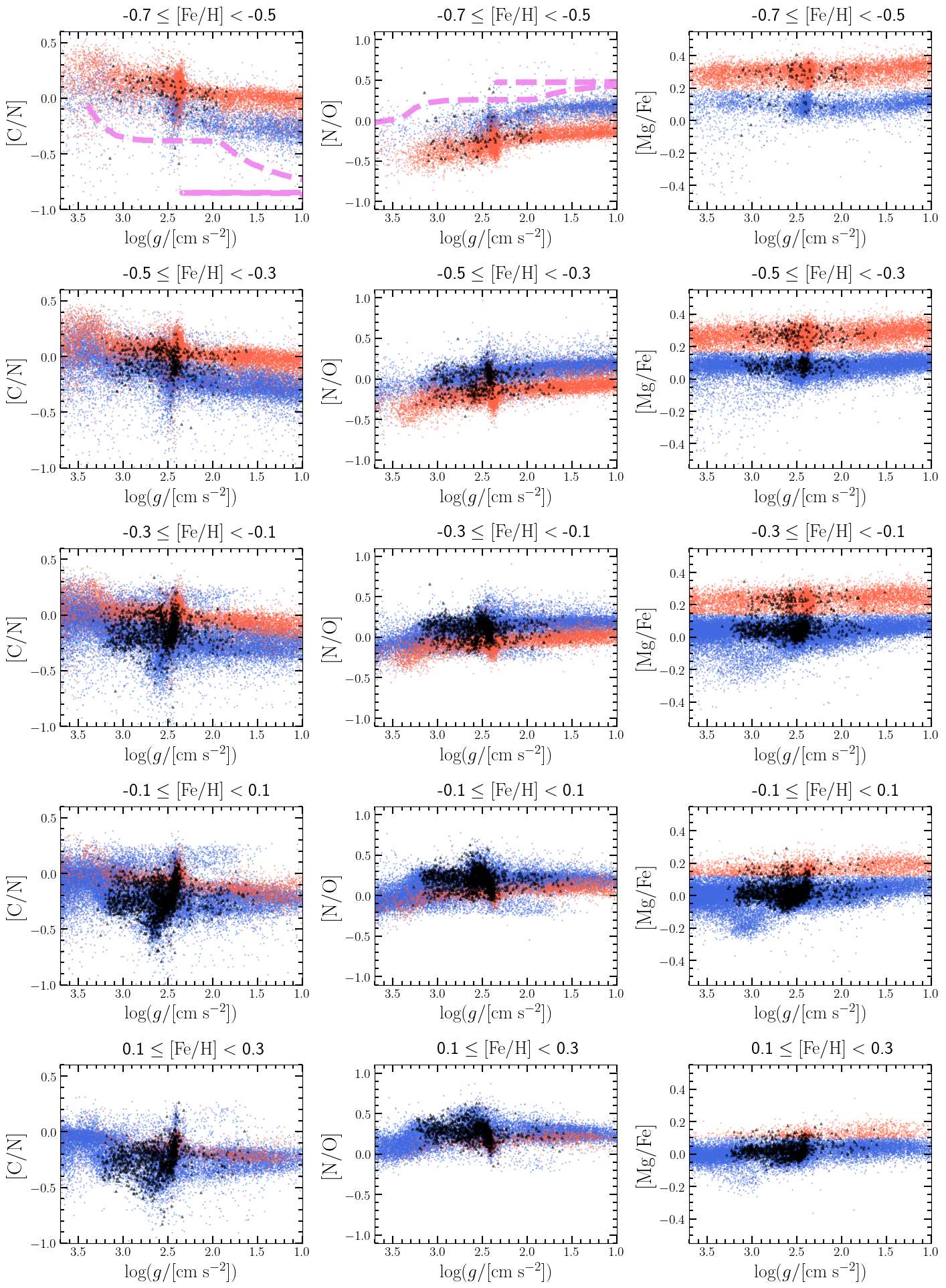}
\caption{The observed [C/N] (first column), [N/O] (second column), and [Mg/Fe] (third column) as a function of $\log(g)$ in the high-$\alpha$ sequence (red points) and low-$\alpha$ sequence (blue points) from APOGEE-DR16, by assuming $\text{SNR} > 80$. Different rows correspond to different ranges of $[\text{Fe/H}]$ abundances. The black triangles correspond to our sample of red giants from \citet{miglio2021}. The magenta dashed lines in the top row show the predictions of the models of \citet{lagarde2012} for a star with initial mass $m=1\,\text{M}_{\sun}$ and metallicity $Z=0.004$ assuming rotation-induced mixing and thermohaline instability. 
}
\label{fig:thick-vs-thin}
\end{figure*}

\subsection{C, N, O surface abundance variations in high-$\alpha$ and low-$\alpha$ stars}

We divide the red giant stars in the sample of \citet{miglio2021} between high-$\alpha$ and low-$\alpha$ by adopting the following demarcation line, which approximately separates them in the $[\text{Mg/Fe}]$ vs $[\text{Fe/H}]$ diagram:
\begin{equation} \label{eq:high-low-alpha}
    \text{[Mg/Fe]} = 0.125 - 0.11\times\text{[Fe/H]}.
\end{equation} 
\noindent For each star in the sample we apply the same analysis as discussed in the previous subsections. We aim at quantifying the surface abundance variations in the high-$\alpha$ thick-disc stars with respect to those in the low-$\alpha$ thin-disc. In Fig. \ref{fig:CNO-high-low-alpha}, we show the results of our analysis for [C/Mg] (first column), [N/Mg] (second column), and [N/O] (third column). In particular, the upper panels show how [C/Mg] vs [Mg/H], [N/Mg] vs [Mg/H], and [N/O] vs [O/H] are observed to differ between the thick-disc (red points) and thin-disc (blue points) stars of our sample. After we apply the surface abundance corrections predicted by our reference MESA stellar evolution models, we obtain surface abundances of [C/Mg] vs [Mg/H], [N/Mg] vs [Mg/H], and [N/O] vs [O/H] shown in the middle panels of Fig. \ref{fig:CNO-high-low-alpha}. Finally, the bottom panels show the model predicted variations of $\log(\text{C/H})$, $\log(\text{N/H})$, and $\log(\text{O/H})$ for thick- and thin-disc stars as a function of [Fe/H]. 

The largest abundance corrections are predicted for thin-disc stars. In particular, the N abundances of thin-disc stars are predicted to be enhanced by as much as $\approx0.45\,\text{dex}$ with respect to the birth abundances. C abundances are predicted to be moderately depleted, with an upper limit in their depletion of the order of  $\approx0.23\,\text{dex}$. We note that the observed [O/Fe] vs [Fe/H] diagram and its bimodal distribution between thick- and thin-disc stars (e.g., see the recent \citealt{vincenzo2021}) are not affected much by mixing processes; the predicted corrections for O are in the range $-0.01\le \Delta\log(\text{O/H})<0.02$, comparable to observational uncertainties. 

After correction, both [C/Mg] and [N/Mg] trends still show a substantial separation between the high-$\alpha$ and low-$\alpha$ populations, indicating that both elements have significant contributions from a ``delayed'' source, most likely AGB enrichment, in addition to prompt enrichment by core-collapse SNe. For [C/Mg], the sequence separation is comparable to that found in GALAH-DR2 by \citet{griffith2019}, but the metallicity dependence is opposite, rising rather than falling with [Mg/H]. For [N/Mg], the strongly rising metallicity dependence for both populations is a sign of metallicity-dependence N yields. In APOGEE data, [O/Mg] ratios are near solar over $-0.7 \la [\text{Mg/H}] \la 0.5$ for both high-$\alpha$ and low-$\alpha$ populations \citep{weinberg2019}, consistent with both elements being produced by core-collapse SNe with IMF-averaged yields that are metallicity independent. Not surprisingly, we find that the trends of [N/O] vs [O/H] are nearly identical to the trends of [N/O] vs [Mg/H]. 

The idea of using C/N ratios as a stellar age diagnostic in APOGEE data \citep{masseron2015,martig2016} is based on the expectation that the surface abundances in APOGEE red giants are driven largely by internal mixing and thus depend strongly on stellar mass. Calibration against asteroseismic masses/ages supports this approach \citep{pinsonneault2018,miglio2021}, as above in Fig. \ref{fig:CN-plots}(a). However, Figs. \ref{fig:CN-plots}(b) and \ref{fig:CNO-high-low-alpha} show that there are also significant dependencies of both C and N abundances on [Fe/H] and on $[\alpha/\text{Fe}]$. These birth abundance trends should be taken into account when using C/N ratios or spectroscopic diagnostics sensitive to them to estimate red giant ages. 

Figure~\ref{fig:thick-vs-thin} places the measurements of our asteroseismic sample in the broader context of APOGEE-DR16 disk CNO measurements.  We plot surface abundances [C/N], [N/O], and [Mg/Fe] against spectroscopic $\log(g)$ over the range $1 \leq \log(g/[\text{cm}\,\text{s}^{-2}]) \leq 3.5$ for stars in the high-$\alpha$ (red) and low-$\alpha$ (blue) populations and for the \cite{miglio2021} stars (larger black dots). Consistent with the findings of \cite{shetrone2019} based on APOGEE-DR13 \citep{holtzmann2018,jonsoon2018}, the [C/N] and [N/O]
ratios of lower metallicity stars ($\text{[Fe/H]} \la -0.5$) show moderate trends
with $\log(g)$ for stars more luminous than the RC (i.e., $\log(g) < 2.5$).
The asteroseismic sample contains too few low metallicity stars to reveal
these trends. \cite{shetrone2019} interpret the decline in [C/N], which
becomes much more prominent at still lower metallicities, as a sign of
extra mixing above the bump in the RGB luminosity function.  For comparison,
in the top row we show the predictions of the \cite{lagarde2012} models
with rotation and thermohaline instability.  At $Z=0.004$ ($\text{[Fe/H]} \approx -0.6$),
these models predict a much stronger $\log(g)$ trend than seen in the
APOGEE observations.  Together with the similarity of RC and RGB ratios
(Figure~\ref{fig:variation-rc-vs-rgb}), this comparison further indicates that extra mixing
in the \cite{lagarde2012} models is too strong, perhaps because of the
numerical or physical treatment of thermohaline instability
(see, e.g., \citealt{constantino2015,constantino2016,constantino2017}).

At low metallicities, the high-$\alpha$ and low-$\alpha$ populations show
clear differences in [C/N] and [N/O] at all $\log(g)$.  The two populations
converge at high metallicity as the separation in [$\alpha$/Fe] itself
becomes smaller.  Figure~\ref{fig:CNO-high-low-alpha} showed similar behavior for the
asteroseismic sample in [C/Mg], [N/Mg], and [N/O], and while correction
for mixing shifts both the high-$\alpha$ and low-$\alpha$ populations in
these diagrams, it does not remove the separation between them.  We conclude
that this separation is, unsurprisingly, a consequence of different birth
abundances.  To interpret this difference, it is useful to recall that
the ``low-$\alpha$'' stars are really Fe rich --- for a given level of
core collapse supernova elements such as O and Mg they have a larger
contribution of Fe from Type Ia supernovae.  The higher [N/O] in this
population implies that N enrichment at least partly tracks this extra
Fe enrichment, probably because it also comes from a time-delayed source
(AGB stars rather than Type Ia supernovae).  The lower [C/N] in this
population implies that C enrichment is not tracking Type Ia Fe to the
same degree.  C behaves partly but not entirely like an $\alpha$ element ---
in Figure~\ref{fig:CNO-high-low-alpha} it shows a separation between low-$\alpha$ and
high-$\alpha$ populations, but only by $\sim 0.1$ dex or less.
This implies a large core collapse supernova contribution to C, in
agreement with theoretical models (e.g.,
\citealt{andrews2017,rybizki2017,griffith2021}).

Figure~\ref{fig:CNO-high-low-alpha} and~\ref{fig:thick-vs-thin} suggest some caution in using C/N ratios or
spectroscopic diagnostics that trace these ratios as age indicators for red
giants.  For stars that are matched in [Fe/H] {\it and} [$\alpha$/Fe], the
differences in surface C/N ratios should mainly reflect differences in the
degree of internal mixing, which in turn depend on mass and thus (for
evolved stars) on age.  However, for stars with different [Fe/H] or different
[$\alpha$/Fe] at the same [Fe/H], differences in surface C/N may also be
affected by differences in birth abundances at the 0.1-0.2 dex level.
One way forward is to use mixing-corrected samples with asteroseismic masses
to define empirical trends of birth abundances (e.g., the middle row of
Figure~\ref{fig:CNO-high-low-alpha}), then use the {\it differences} from these expected
birth abundances, $\Delta$[C/N] and $\Delta$[N/O], as the diagnostic for
ages in stars that do not have asteroseismic masses.

\begin{figure}    
\centering
\includegraphics[width=8cm]{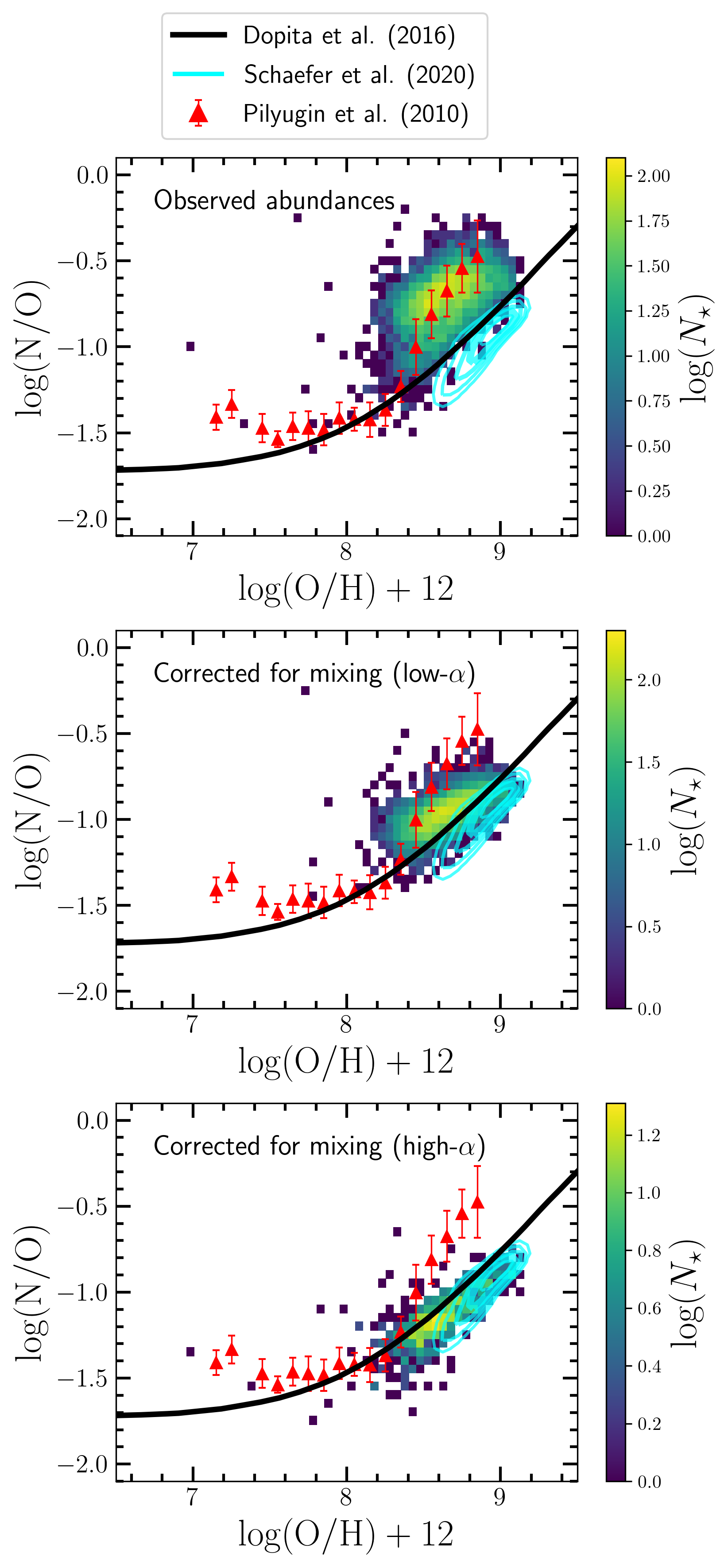}
\caption{Comparison between $\log(\text{N/O})$ vs $\log(\text{O/H})+12$ as observed in the sample of \citet[2-D histograms in both panels]{miglio2021} and observations in resolved HII regions of nearby galaxies \citep[``ONS'' calibration]{pilyugin2010} and unresolved star-forming regions of external galaxies \citet[``R23'' calibration of \citealt{maiolino2008}]{schaefer2020}. The black solid line corresponds to the metallicity calibration relation as proposed by \citet{dopita2016}. The 2-D histogram in the top panel shows the observed APOGEE-DR16 abundances of the red giants in \citet{miglio2021}. The 2-D histograms in the middle and bottom panels show the results of our analysis for stars in the low- and high-[Mg/Fe] sequences, respectively (see equation \ref{eq:high-low-alpha}), when we correct the observed abundances of O and N for mixing processes. 
}
\label{fig:comparison-gas-phase}
\end{figure}

\section{Comparison with gas-phase abundances}
\label{sec:comparing}

Oxygen abundance measurements in HII regions from strong nebular lines are affected by large systematic uncertainty. Such measurements usually rely on indexes and calibration relations based on sub-samples where weak auroral lines are available for the so-called direct or $T_{\text{e}}$-method to be applied. Uncertainties in the derived O abundances can be as large as $\approx0.6\,\text{dex}$ (e.g., \citealt{kewley2008,moustakas2010,lopez-sanchez2012,blanc2015,belfiore2017}), and widely used calibrations rely on indexes based on $\log(\text{N/O})$ (see the discussion and references in \citealt{schaefer2020}). 

Since the average $\log(\text{N/O})$ vs $\log(\text{O/H})+12$ relation is mostly determined by the nucleosynthesis of N, which strongly depends on metallicity, the surface N and O abundances in red giants corrected for the mixing processes can be used to constrain the best calibration relation to measure $\log(\text{O/H})+12$ and $\log(\text{N/O})$ in the HII regions of our Galaxy and external galaxies in the high-metallicity regime where weak auroral lines are not available.

The results of our analysis are shown in Fig. \ref{fig:comparison-gas-phase}. In the bottom panel, we compare $\log(\text{N/O})$ vs $\log(\text{O/H})+12$ as observed in the red giants of \citet[2-D histogram]{miglio2021} with the N and O abundance measurements in a sample of resolved HII regions in extragalactic systems \citep[``ONS'' calibration]{pilyugin2010}; we also show the abundance measurements in the star-forming regions of large samples of galaxies with different stellar mass by \citet[``R23'' calibration of \citealt{maiolino2008}]{schaefer2020}, and the calibration relation proposed by \citet{dopita2016}, which was also used as an observational constraint  by \citet{vincenzo2018a,vincenzo2018b} for a comparison with the predictions of their cosmological simulation for N and O abundances in galaxies. 

If our reference MESA stellar evolution models are able to correctly characterize the surface abundance variations of N and O in red giants of different mass and metallicity, then the best agreement with the predicted surface abundance measurements corrected for mixing is obtained with the calibration relation proposed by \citet{dopita2016}. Interestingly, when we apply the surface abundance correction, we can obtain a better agreement also with the observations of \citet{schaefer2020} for MaNGA galaxies, who assumed the ``R23'' calibration of \citet{maiolino2008}, whereas the ``ONS'' calibration of \citet{pilyugin2010} seems to be disfavoured. A similar conclusion was found in the analysis of \citet{magrini2018}, who presented MW stellar abundances of N and O as measured by the Gaia-ESO spectroscopic survey. 
In detail, our results for low-$\alpha$ stars lie slightly above the \citet{dopita2016} calibration (by about $0.1\,\text{dex}$), while our results for high-$\alpha$ stars is slightly below the \citet{dopita2016} calibration (by about $0.05\,\text{dex}$) and agree well with the MaNGA results. 

\begin{figure}    
\centering
\includegraphics[width=8cm]{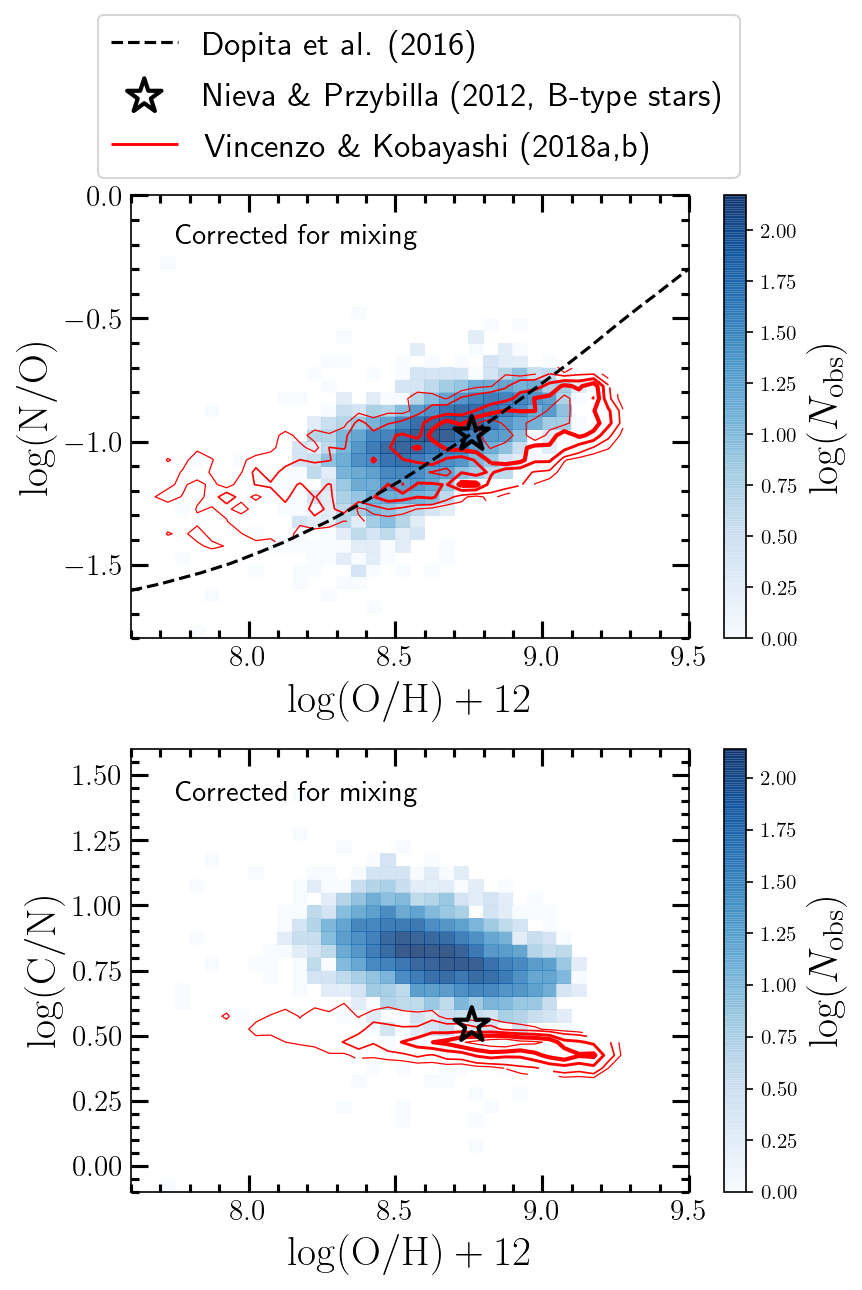}
\caption{Comparison between the evolution-corrected $\log(\text{N/O})$ vs $\log(\text{O/H})+12$ (top panel) and $\log(\text{C/N})$ vs $\log(\text{O/H})+12$ (bottom panel) in the sample of \citet[2-D histograms in both panels]{miglio2021} with the predicted stellar abundances in Galaxy B of \citet[red solid contours corresponding to the normalized number densities of $10$, $20$, $40$, $60$, and $80$ per cent]{vincenzomiglio2019} from the cosmological hydrodynamical simulation of \citet{vincenzo2018a,vincenzo2018b}. The black dashed line corresponds to the metallicity calibration relation as proposed by \citet{dopita2016}. The black star symbols show the average abundances in a sample of B-type stars in the Solar neighbourhood as measured by \citet{nieva2012}. 
}
\label{fig:comparison-simulation}
\end{figure}

In Fig. \ref{fig:comparison-simulation}, we compare the evolution-corrected $\log(\text{N/O})$ vs $\log(\text{O/H})+12$ (top panel) and $\log(\text{C/N})$ vs $\log(\text{O/H})+12$ (bottom panel) in our sample of red giants with the predicted stellar abundances in a simulated star-forming disc galaxy selected in the cosmological hydrodynamical simulation of \citet{vincenzo2018a,vincenzo2018b}, corresponding to Galaxy B in \citet{vincenzomiglio2019}. The agreement is excellent for $\log(\text{N/O})$ vs $\log(\text{O/H})+12$, but there is a strong tension for $\log(\text{C/N})$ between cosmological simulations and the results our analysis based on the MESA models. The simplest interpretation of this discrepancy is that the C yields adopted in these simulations are too high. 

\section{Conclusions}
\label{sec:conclusions}

In this paper, we have studied how $\log(\text{C/N})$ and $\log(\text{N/O})$ evolve in a sample of red giants as functions of their mass, metallicity, and evolutionary phase, by using chemical abundance measurements from the APOGEE collaboration (\citealt{majewski2017}; DR16 public release of \citealt{ahumada2020}) and precise masses and radii from the asteroseismic analysis of \citet{miglio2021}, which was performed on the light curves of the stars as observed the \textit{Kepler} mission (\citealt{borucki2010}; see Section \ref{sec:miglio-sample} for more details). 

Observational data have been compared with the predictions of our reference set of stellar evolution models developed with the code {\tt MESA} \citep{paxton2011,paxton2013,paxton2015,paxton2018} by assuming standard mixing prescriptions for a fine grid of stellar masses and [Fe/H] abundances, with $[\alpha/\text{Fe}]=0.0$, $0.2$, and $0.4$ (see Section \ref{sec:MESA-models} for more details). We have also analysed an alternative set of stellar evolution models developed by \citet{lagarde2012} with the code {\tt STAREVOL} \citep{siess2000,palacios2003,palacios2006,decressin2009} for a relatively coarse grid of masses and metallicities, by assuming either standard mixing prescriptions or extra-mixing mechanisms (rotation-induced mixing and thermohaline instability; see Section \ref{sec:lagarde-models} for more details). 

Our conclusions can be summarized as follows.
\begin{enumerate}
    \item By selecting the stars classified as RGB in the analysis of \citet{miglio2021}, the observed $\log(\text{N/O})$ ratios are increasingly enhanced at the stellar surface  when moving to larger masses at fixed metallicity and surface gravity. Conversely, the observed $\log(\text{C/N})$ ratios are increasingly depleted at the surface. Our reference models with standard mixing prescriptions can qualitatively reproduce these trends. As expected, models with extra-mixing predict larger $|\Delta\log(\text{N/O})|$ and $|\Delta\log(\text{C/N})|$ at fixed stellar masses than models with standard mixing prescriptions, but the predicted behaviour of $\log(\text{N/O})$ and $\log(\text{C/N})$ vs stellar mass is similar in both cases. Over the surface gravity range probed by our sample, $1.5 \la \log(g/[\text{cm}\,\text{s}^{-2}])\la 3.3$, the trends of observed abundances are nearly flat in $\log(g)$ at fixed stellar mass, in agreement with the predictions of the models.
    
    \item The slope of $\log(\text{N/O})$ and $\log(\text{C/N})$ as a function of stellar mass is observed to change at $m \approx 1.2\,\text{M}_{\sun}$, which approximately divides stars with a radiative core in the MS from stars with a convective core.  

    \item The stellar models of \citet{lagarde2012} with rotation-induced mixing and thermohaline instability predict RC stars to have higher $\log(\text{N/O})$ and lower $\log(\text{C/N})$ ratios than RGB stars of the same mass and metallicity. In the observed sample, RGB and RC stars are characterized by similar median values of $\log(\text{N/O})$ and $\log(\text{C/N})$ as a function of stellar mass. If we also look at the tails of the observed distributions, then the disagreement between the extra-mixing model predictions and observations is particularly evident for $\log(\text{C/N})$. 
    
    \item We develop a formalism to calculate $\log(\text{C/H})$, $\log(\text{N/H})$, and $\log(\text{O/H})$ that one would get with different birth abundances from those used in the MESA models. Differences of the order of $\delta\text{C}_{\sun}=0.11$, $\delta\text{N}_{\sun}=-0.07$, and $\delta\text{O}_{\sun}=0.07$ relative to \citet{asplund2009}, together with a metallicity trend $\delta(\text{N}) \propto 0.7 \times [\text{Fe/H}]$, can reproduce the observed behaviour of $\log(\text{N/O})$ and $\log(\text{C/N})$ vs mass in different metallicity ranges for our MESA standard mixing models (see Fig. \ref{fig:CN-NO-corrections}). The quantities $\delta\text{N}_{\sun}$ and $\delta\text{O}_{\sun}$ are consistent with the difference between \citet{nieva2012} for a sample of B-type stars in the Solar neighbourhood and \citet{asplund2009} for the solar photosphere; conversely, $\delta\text{C}_{\sun}$ differs by $0.2\,\text{dex}$ with respect to the difference between \citet{nieva2012} and \citet{asplund2009} (which is $-0.1\,\text{dex}$). The assumed metallicity trend for the birth abundances of N is based on widely used metallicity calibrations to measure gas-phase abundances from strong nebular emission lines (see Fig. \ref{fig:comparison-gas-phase}). 

    \item When moving to younger ages, red giants are observed to have increasingly high $\log(\text{N/O})$ and low $\log(\text{C/N})$ ratios at fixed [Fe/H] abundance. If  we fix the stellar age, $\log(\text{N/O})$ and $\log(\text{C/N})$ correlate with [Fe/H], such that more metal-rich stars have higher $\log(\text{N/O})$ and lower $\log(\text{C/N})$ ratios. When we apply the abundance corrections due to mixing as predicted by our reference stellar evolution models, the observed age-dependence of $\log(\text{N/O})$ and $\log(\text{C/N})$ is effectively removed. Since the predicted abundance corrections weakly depend on metallicity, after we correct the observed abundances for the mixing evolutionary processes, we are left with a residual dependence of $\log(\text{N/O})$ and $\log(\text{C/N})$ on [Fe/H], which is particularly strong for $\log(\text{N/O})$, as expected by the nucleosynthesis studies of C and N in galaxies (e.g., see \citealt{vincenzo2016,vincenzo2018a}, and references therein). 
    
    \item The largest abundance corrections are predicted for thin-disc stars, i.e., stars with approximately solar $[\alpha/\text{Fe}]$ ratios. The [N/H] ratios of thin-disc stars are predicted to be enhanced by up to  $0.45\,\text{dex}$ with respect to the birth abundances. [C/H] ratios are predicted to be moderately depleted by up to $0.23\,\text{dex}$. The observed [O/Fe] vs [Fe/H] diagram and its bimodal distribution between thick- and thin-disc stars is not affected by mixing processes; the predicted corrections for O are in the range $-0.01\le \Delta\log(\text{O/H})<0.02$.
    
    \item Thick-disc (high [$\alpha$/Fe]) stars are observed to have systematically higher [C/N] and lower [N/O] than thin-disc stars with the same metallicity and surface gravity. The difference persists after mixing corrections, so it is likely due to different C, N, and O birth abundances between thick- and thin-disc stars. This difference can have a significant impact on spectroscopic age determinations based on C/N if ignored. The trend of [C/N] with [Fe/H] could also affect C/N-based age determinations.

    \item Our asteroseismic sample does not have enough low metallicity stars to reveal the dependence of observed [C/N] and [N/O] on $\log(g)$ found by \cite{shetrone2019} for low metallicity stars in APOGEE-DR13 (see also \citealt{masseron2017}).  We do see moderate trends with $\log(g)$ for $-0.7 \leq [\text{Fe/H}] \leq -0.5$ when considering the full APOGEE-DR16 disk sample (Fig.~\ref{fig:thick-vs-thin}), consistent with the trend found by \cite{shetrone2019} at this metallicity.  For $Z=0.004$ ($\text{[Fe/H]} \approx -0.6$), the \cite{lagarde2012} models with rotation and thermohaline instability predict a change in [C/N] at $\log(g) < 2.0$ that is much larger than observed.   
    
    \item The predicted $\log(\text{N/O})$ vs $\log(\text{O/H})+12$ abundance pattern as corrected for mixing in our sample of red giants is in excellent agreement with the gas-phase calibration relation proposed by \citet{dopita2016} to measure chemical abundances at high-redshift. A good agreement is also obtained with the abundance measurements of \citet{schaefer2020} in MaNGA galaxies, who assumed the ``R23'' calibration of \citet{maiolino2008}, whereas the ``ONS'' calibration of \citet{pilyugin2010} seems to be disfavoured. 
    
    \item The evolution-corrected $\log(\text{N/O})$ vs $\log(\text{O/H})+12$ in our sample of red giants is in excellent agreement with the predictions in the simulated star-forming disc galaxies from the cosmological simulation of \citet{vincenzo2018a,vincenzo2018b,vincenzomiglio2019}. The simulations do not agree with the mixing-corrected $\log(\text{C/N})$ ratios, a discrepancy that most likely reflects incorrect C yields.
    
\end{enumerate}

Our ability to stringently test mixing models is limited by uncertainties in
the appropriate choice of birth abundances.  However, based on
Fig.~\ref{fig:CN-NO-corrections} we conclude that our MESA models with standard mixing
are consistent with APOGEE+{\it Kepler} data to within plausible uncertainties
in the birth abundances and calibration uncertainties of the C, N, and O
measurements.  The mixing-corrected [C/N], and [N/O] trends in
Figs.~\ref{fig:NO-plots} and~\ref{fig:CN-plots} and [C/Mg] and [N/Mg] trends in
Fig.~\ref{fig:CNO-high-low-alpha} therefore provide a reasonable guide to chemical evolution
of those elements in the MW disk.  The separation of low-$\alpha$ and
high-$\alpha$ populations implies that N and (to a lesser degree) C have
significant contributions from time-delayed sources, and the [Fe/H]
trends imply metallicity dependence of N yields.  These empirical trends
allow new quantitative tests of stellar yield calculations and chemical
evolution scenarios.


\section{Acknowledgments} 

We thank Nad\`{e}ge Lagarde for useful suggestions, and Adam Schaefer for kindly providing the observational data for MaNGA galaxies shown in Fig. 11. 
This work was supported in part by NSF grant AST-1909841.
FV acknowledges the support of a Fellowship from the Center for Cosmology and AstroParticle Physics at the Ohio State University. DW acknowledges the hospitality of the Institute for Advanced Study and the support of the W.M. Keck Foundation.  JM, AM, and SK acknowledge support from the ERC Consolidator Grant funding scheme (project ASTEROCHRONOMETRY, G.A. n. 772293). SH is supported by an NSF Astronomy and Astrophysics Postdoctoral Fellowship under award AST-1801940.

This research has made use of the VizieR catalogue access tool, CDS,
 Strasbourg, France (DOI : 10.26093/cds/vizier; website: \url{http://vizier.unistra.fr}). The original description of the VizieR service was published in \citet{ochsenbein2000}.
 
 This paper includes data collected by the \textit{Kepler} mission and obtained from the MAST data archive at the Space Telescope Science Institute (STScI). Funding for the \textit{Kepler} mission is provided by the NASA Science Mission Directorate. STScI is operated by the Association of Universities for Research in Astronomy, Inc., under NASA contract NAS 5–26555.
 
In this work we have made use of SDSS-IV APOGEE-2 DR16 data. Funding for the Sloan Digital Sky Survey IV has been provided by the Alfred P. Sloan Foundation, the U.S. Department of Energy Office of Science, and the Participating Institutions. SDSS-IV acknowledges
support and resources from the Center for High-Performance Computing at
the University of Utah. The SDSS web site is \url{www.sdss.org}.

SDSS-IV is managed by the Astrophysical Research Consortium for the 
Participating Institutions of the SDSS Collaboration including the 
Brazilian Participation Group, the Carnegie Institution for Science, 
Carnegie Mellon University, the Chilean Participation Group, the French Participation Group, Harvard-Smithsonian Center for Astrophysics, 
Instituto de Astrof\'isica de Canarias, The Johns Hopkins University, Kavli Institute for the Physics and Mathematics of the Universe (IPMU) / 
University of Tokyo, the Korean Participation Group, Lawrence Berkeley National Laboratory, 
Leibniz Institut f\"ur Astrophysik Potsdam (AIP),  
Max-Planck-Institut f\"ur Astronomie (MPIA Heidelberg), 
Max-Planck-Institut f\"ur Astrophysik (MPA Garching), 
Max-Planck-Institut f\"ur Extraterrestrische Physik (MPE), 
National Astronomical Observatories of China, New Mexico State University, 
New York University, University of Notre Dame, 
Observat\'ario Nacional / MCTI, The Ohio State University, 
Pennsylvania State University, Shanghai Astronomical Observatory, 
United Kingdom Participation Group,
Universidad Nacional Aut\'onoma de M\'exico, University of Arizona, 
University of Colorado Boulder, University of Oxford, University of Portsmouth, 
University of Utah, University of Virginia, University of Washington, University of Wisconsin, 
Vanderbilt University, and Yale University.

\section*{Data availability}
The data underlying this article, including abundances corrected for evolutionary mixing, will be shared on reasonable request to the corresponding author.

\end{document}